\documentclass[twocolumn,trackchanges]{aastex701}

\usepackage{soul}
\usepackage{amsmath}	
\usepackage{amssymb}	
\usepackage{enumitem}
\usepackage{MnSymbol}
\usepackage{hyperref}
\usepackage{verbatim} 
\usepackage{array}
\usepackage[section]{placeins} %
\usepackage{graphicx}
\usepackage{caption}
\usepackage{subcaption}
\usepackage{placeins}
\usepackage{booktabs}
\newcommand{\change}[1]{\textcolor{black}{#1}}
\newcommand{\changeblue}[1]{\textcolor{black}{#1}}
\newcommand{\changegreen}[1]{\textcolor{black}{#1}}
\usepackage{acro}
\DeclareAcronym{tess}{
  short = {TESS},
  long = {Transiting Exoplanet Survey Satellite}
}
\DeclareAcronym{TIC}{
  short = {TIC},
  long = {TESS Input Catalog}
}

\DeclareAcronym{MAST}{
  short = {MAST},
  long = {Mikulski Archive for Space Telescopes}
}

\DeclareAcronym{SPOC}{
  short = {SPOC},
  long = {Science Processing Operations Center} 
}

\DeclareAcronym{SPC}{
  short = {SPC},
  long = Stellar parameter classification
}

\acsetup{patch/longtable=false}

\shortauthors{Redyan et al.}
\received{September 23, 2025}
\revised{May 23, 2026 }
\accepted{May 23, 2026}
\submitjournal{AJ}

\begin{document}

\title{TOI-2155\,b: A Massive Brown Dwarf or a Very Low-Mass Star?}

\author[0000-0001-7483-9982]{Md Redyan Ahmed}
\affiliation{Sydney Institute for Astronomy, School of Physics, University of Sydney, NSW 2006,
Australia}
\email{mdredyan.ahmed@sydney.edu.au}

\author[0000-0002-6939-9211]{Tansu Daylan}
\affiliation{Department of Physics and McDonnell Center for the Space Sciences, Washington University, St. Louis, MO 63130, USA}
\email{email_here}

\author[0000-0001-6416-1274]{Theron W. Carmichael}
\affiliation{Institute for Astronomy, University of Hawaii at Manoa, 2680 Woodlawn Drive, Honolulu, HI 96822, USA}
\email{email_here}

\author[0000-0003-2478-0120]{Sarah L Casewell }
\affiliation{ School of Physics and Astronomy, University of Leicester, University Road, Leicester, LE1 7RH, UK}
\email{email_here}

\author[orcid=0000-0003-0353-9741,gname=Anita,sname=Hafner]{Anita Hafner}
\affiliation{Sydney Institute for Astronomy, School of Physics, University of Sydney, NSW 2006, Australia}
\email{anita.hafner@sydney.edu.au}

\author[orcid=0000-0003-0353-9741,gname=Jaime A.,sname=Alvarado-Montes]{Jaime A. Alvarado-Montes}
\altaffiliation{Macquarie University Research Fellow (MQRF)}
\affiliation{Australian Astronomical Optics, Macquarie University, Balaclava Road, Sydney, NSW 2109, Australia.}
\affiliation{Astrophysics and Space Technologies Research Centre, Macquarie University, Balaclava Road, Sydney, NSW 2109, Australia.}
\email{jaime.alvaradomontes@mq.edu.au}

\author[0000-0001-6637-5401]{Allyson~Bieryla} 
\affiliation{Center for Astrophysics ${\rm \mid}$ Harvard {\rm \&} Smithsonian, 60 Garden Street, Cambridge, MA 02138, USA}
\email{email_here}
\author[0000-0002-8964-8377]{Samuel N.\ Quinn}
\affiliation{Center for Astrophysics ${\rm \mid}$ Harvard {\rm \&} Smithsonian, 60 Garden Street, Cambridge, MA 02138, USA}
\email{squinn@cfa.harvard.edu}

\author[0000-0002-2830-5661]{Michael Calkins}
\affiliation{Center for Astrophysics ${\rm \mid}$ Harvard {\rm \&} Smithsonian, 60 Garden Street, Cambridge, MA 02138, USA}
\email{email_here}
\author[0000-0001-6588-9574]{Karen A.\ Collins}
\affiliation{Center for Astrophysics ${\rm \mid}$ Harvard {\rm \&} Smithsonian, 60 Garden Street, Cambridge, MA 02138, USA}
\email{karen.collins@cfa.harvard.edu }

\author[0000-0001-8621-6731]{Cristilyn N.\ Watkins}
\affiliation{Center for Astrophysics ${\rm \mid}$ Harvard {\rm \&} Smithsonian, 60 Garden Street, Cambridge, MA 02138, USA}
\email{cristilyn.watkins@cfa.harvard.edu}

\author[0000-0002-3481-9052]{Keivan G.\ Stassun}
\affiliation{Department of Physics and Astronomy, Vanderbilt University, Nashville, TN 37235, USA}
\email{email_here}

\author[0000-0003-1713-3208]{Boris S. Safonov}
\affiliation{Sternberg Astronomical Institute, M.V. Lomonosov Moscow State University, 13, Universitetskij pr., 119234, Moscow, Russia.}
\email{email_here}

\author[0000-0003-2228-7914]{Maria V. Goliguzova}
\affiliation{Sternberg Astronomical Institute, M.V. Lomonosov Moscow State University, 13, Universitetskij pr., 119234, Moscow, Russia.}
\email{email_here}

\author[0000-0001-8134-0389]{Giuseppe Marino} 
\affiliation{Wild Boar Remote Observatory, San Casciano in val di Pesa, Firenze, 50026 Italy}
\affiliation{INAF - Osservatorio Astrofisico di Catania, Via S. Sofia 78, 95123 Catania, Italy}
\email{giumar69@gmail.com}

\author[0000-0003-2239-0567]{Dennis M.\ Conti} 
\affiliation{American Association of Variable Star Observers, 185 Alewife Brook Parkway, Suite 410, Cambridge, MA 02138, USA}
\email{dennis@astrodennis.com}


\author[0000-0001-7026-6291]{Peter Tuthill}
\affiliation{Sydney Institute for Astronomy,School of Physics, University of Sydney, NSW 2006,
Australia}
\email{peter.tuthill@sydney.edu.au}




\begin{abstract}

We present TOI-2155\,b, \change{a massive transiting companion}, discovered using data from NASA's Transiting Exoplanet Survey Satellite (TESS) mission and confirmed with ground-based RV measurements from the Tillinghast Reflector Echelle Spectrograph (TRES). We also analyze ground-based follow-up photometric data from the Wendelstein Observatory (WST), Las Cumbres Observatory Global Telescope (LCOGT), and Wild Boar Remote Observatory (WBR). TOI-2155\,b is a short-period companion with {$P= 3.7246950 \pm{0.0000014}$}~days.  The radius and mass of TOI-2155\,b are found to be $R_b = 0.972^{+0.009}_{-0.008} \,\mathrm{R_J}$ and $M_b = 80.6^{+1.0}_{-1.1} \,\mathrm{M_J}$, respectively, corresponding to a density of {$\rho_b= 109^{+3.1}_{-3.3}$ g cm$^{-3}$}. The F-type subgiant host star has an effective temperature of $T_{\rm eff} = 6085\pm 78$ K, a radius $R_{\thinstar} = 1.705^{+0.066}_{-0.064}$ $\mathrm{R_\odot}$ and a mass $M_\star = 1.33 \pm 0.008$~M$_\odot$. 
With a mass close to the hydrogen-burning minimum mass, TOI-2155\,b lies at the boundary between brown dwarfs and low-mass stars. Its measured mass, radius, and density place it in a transitional region, where distinguishing between a massive brown dwarf and a very low-mass star is not straightforward. TOI-2155\,b therefore provides a valuable benchmark for testing evolutionary models of stellar and substellar structure near the hydrogen-burning limit.

\end{abstract}


\section{Introduction}

The search for exoplanets in recent decades has also deepened our understanding of brown dwarfs (BDs): substellar objects that occupy the region between planets and stars, \changeblue{generally defined as having masses $\sim$13-80 M$_{\mathrm{J}}$, with} radii ranging from 0.7 to 1.4  R$_{\mathrm{J}}$ \citep{vsubjak2020toi, Csizmadia_2016, deleuil2008transiting, RevModPhys.73.719}. However, a simple mass-based definition is not enough to classify BDs. These traditional boundaries of BDs are approximate and can be affected by metallicity, \change{helium abundance, age, initial deuterium abundance,} and other factors, \changeblue{with the lower limit ranging from} 11–16 M$_{\mathrm{J}}$ \citep{spiegel2011deuterium} and \changeblue{the upper limit ranging from} 75–80 M$_{\mathrm{J}}$ \citep{baraffe2002evolutionary}. 

\changeblue{The atmosphere models of \citet{morley2024sonora} show that the hydrogen-burning minimum mass can be shifted further upward under specific conditions, such as low metallicity and cloud-free atmospheres, reaching $\sim81.7\,M_{\rm J}$, while ATMO models for solar-metallicity objects place the hydrogen-burning minimum mass (HBMM) at $\sim78.5\,M_{\rm J}$ \citep{chabrier2023impact}. This implies that the mass boundary between brown dwarfs and low-mass stars is not universal, making companions near this transition region particularly interesting.}

\change{Brown dwarfs are supported primarily by electron degeneracy pressure. \changeblue{Observational studies suggest that the mass–density relation of substellar objects reaches a maximum in the brown dwarf regime. \citet{hatzes2015definition} identified a density peak around $\sim60\,M_{\rm J}$, while \citet{persson2019greening} suggested that it may lie closer to $\sim73\,M_{\rm J}$ which corresponds to the mass range close to the hydrogen-burning minimum mass.}}

\change{Objects with masses close to the hydrogen-burning minimum mass occupy a transitional regime between brown dwarfs and hydrogen-burning stars. In this mass range, companions may either sustain stable proton--proton fusion or cool and contract as brown dwarfs, depending on their interior properties and composition. Objects in this mass range are particularly useful for testing evolutionary models of both brown dwarfs and very low-mass stars. \changeblue{Precise determination of their masses and radii allow comparison with model predictions of the hydrogen-burning limit, although masses and radii do not by themselves directly constrain the hydrogen-burning limit, as it remains model-dependent and sensitive to assumptions such as metallicity, the equation of state, and atmospheric/cloud treatment.}}

\changeblue{Although brown-dwarf formation pathway is still debated, BDs and low-mass stars likely form through similar processes, including gravitational collapse of molecular cloud cores and  disc fragmentation}. However, unlike low-mass stars, BDs do not sustain hydrogen fusion and therefore evolve differently over time \citep{Chabrier_2000}. \changeblue{Brown dwarfs may undergo brief deuterium- or lithium-burning phases, and the most massive ones near the hydrogen-burning limit may briefly burn some hydrogen, but they do not sustain stable hydrogen fusion \citep{d1997evolution,myers1983dense,beichman1986candidate,larson1981turbulence}.}

\changeblue{Precise measurements of the masses and radii (and hence densities) of stellar and substellar companions allow comparison with theoretical models for brown dwarfs and very low-mass stars \citep[e.g.,][]{saumon2008evolution,baraffe2003evolutionary,chabrier1997structure}. These measurements are best achieved for transiting systems, where precise light curves from the Transiting Exoplanet Survey Satellite (TESS) constrain key orbital parameters such as the period and inclination, as well as the radius of the transiting companion. When combined with radial velocity (RV) observations, they allow us to determine companion masses and bulk densities \citep{henderson2024ngts,carmichael2022toi}}.

While large-scale surveys have identified over 1000 solitary BDs in the Solar neighborhood \citep{burningham2018}, only $\sim$50 transiting brown dwarf systems have been discovered to date. In recent years, TESS has been a primary tool for detecting new transiting BDs \citep[e.g.,][]{vowell202511}, detecting more than two-thirds of these 50 systems \citep[e.g.][and references therein]{zhang2026oasis}.

\change{In this paper, we introduce TOI-2155\,b, a transiting companion with a mass near the hydrogen-burning minimum mass limit, placing it close to the boundary between brown dwarfs and very low-mass stars}. We describe the observations collected for TOI-2155 in Section~\ref{sec: observations}. The properties of the host star as well as the modelling of TOI-2155\,b are explained in Section~\ref{sec: results}. In \change{Section~\ref{sec: Discussion}, we compare TOI-2155\,b to known transiting brown dwarfs and very low-mass stellar companions, placing it in the broader context of objects near the brown dwarf– low mass star boundary}. Finally, we conclude in Section~\ref{sec:conclusion} with a summary of this work.

\section{Observations} 
\label{sec: observations}

\begin{figure*}
   \centering
\includegraphics[width=1\textwidth, height=0.9\textheight, keepaspectratio]{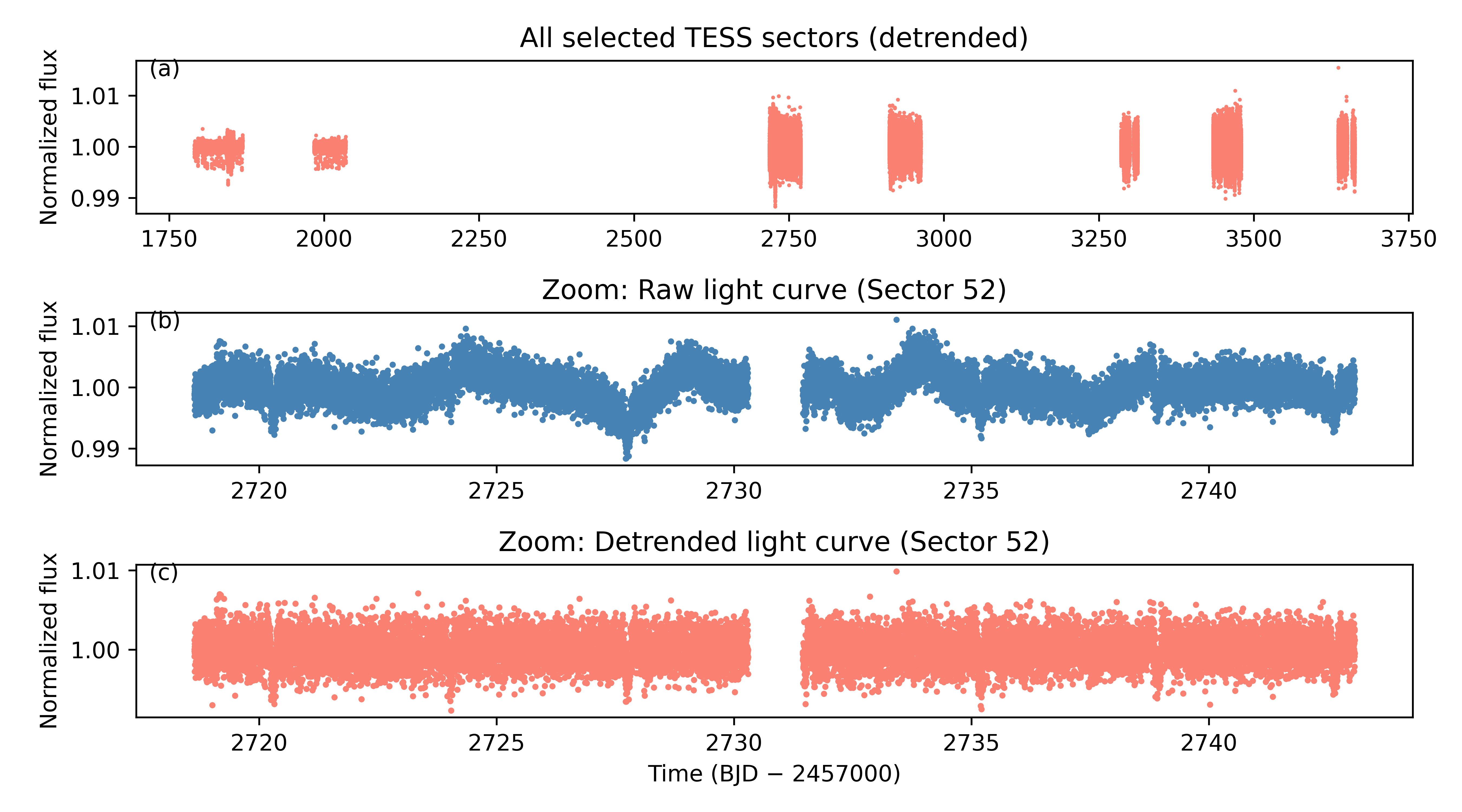}
   \caption{\change{Data of TOI-2155\,b from 13 TESS sectors. Panel (a) shows the detrended light curve from all TESS sectors using the biweight filter implemented in wōtan with a window length of 0.5 days. Panel (b) presents a zoomed view of the raw PDCSAP flux from Sector 52. Panel (c) shows the detrended data of Sector 52. The detrended light curve was used in the global modeling of the TOI-2155 system.}}
   \label{fig: detrended_flux}
\end{figure*}

A summary of the facilities used in this work is provided in Table \ref{facilities}. 

\begin{table*}
\centering
\caption{This table serves as a reference for the facilities used to confirm and analyze the TOI-2155b.}
\label{facilities}
\begin{tabular}{|l|l|}
\hline
\textbf{Facility} & \textbf{Data Type} \\
\hline
Transiting Exoplanet Survey Satellite (TESS) & Transit photometry \\
Wendelstein Observatory (WST) & Transit photometry  \\
Wild Boar Remote Observatory (WBR) & Transit photometry  \\
Las Cumbres Observatory Global Telescope (LCOGT)  & Transit photometry \\
Caucasian Mountain Observatory (CMO) & Speckle imaging \\
 Tillinghast Reflector Echelle Spectrograph (TRES) & Spectroscopy \& Radial velocity \\
\hline
\end{tabular}
\end{table*}

\subsection{TESS light curves}
 
TOI-2155 has a \ac{TIC} ID 461591646. We use publicly accessible data from the \ac{MAST}, supplied by \ac{SPOC}. The common systematics are eliminated from the Presearch Data Conditioned Simple Aperture Photometry (PDCSAP) light curves. The light curve and related uncertainties were extracted from the original scientific data using the SPOC pipeline \citep{jenkins2016tess}. To detrend data and remove long-term systematics (Figure \ref{fig: detrended_flux}), we use a biweight algorithm with a window length of 0.5 days as implemented in wōtan \citep{Hippke_2019}. Table~\ref{tess_sector_details_all} lists the TESS sectors in which TOI-2155\,b was observed, along with the number of transits and observation dates.


\begin{table*}
\centering
\caption{Photometric observations of TOI-2155\,b by TESS across different sectors.}
\label{tess_sector_details_all}
\hspace{-25pt}
\begin{tabular}{|c|c|c|c|c|}
\hline 
\label{tess_sector_details_all}
\textbf{Sector} & \textbf{Transits} & \textbf{Observation date} & \textbf{Pipeline} & \textbf{Cadence (s)} \\
\hline
\change{18} & \change{6} & \change{2019 Nov 02 to 2019 Nov 27} & \change{TESS-SPOC} & \change{1800} \\
\change{19} & \change{7} & \change{2019 Nov 28 to 2019 Dec 23} & \change{TESS-SPOC} & \change{1800} \\
\change{20} & \change{7} & \change{2019 Dec 24 to 2020 Jan 20} & \change{TESS-SPOC} & \change{1800} \\
\change{25} & \change{7} & \change{2020 May 14 to 2020 Jun 08} & \change{TESS-SPOC} & \change{1800} \\
\change{26} & \change{6} & \change{2020 Jun 09 to 2020 Jul 04} & \change{TESS-SPOC} & \change{1800} \\

\hline
52 & 7 & 2022 May 19 to 2022 Jun 12 & SPOC & \change{120} \\
53 & 7 & 2022 Jun 13 to 2022 Jul 08 & SPOC & \change{120} \\
59 & 7 & 2022 Nov 26 to 2022 Dec 23 & SPOC & \change{120} \\
60 & 6 & 2022 Dec 23 to 2023 Jan 18 & SPOC & \change{120} \\
73 & 6 & 2023 Dec 07 to 2024 Jan 03 & SPOC & \change{120} \\
78 & 5 & 2024 May 03 to 2024 May 21 & SPOC & \change{120} \\
79 & 7 & 2024 May 22 to 2024 Jun 18 & SPOC & \change{120} \\
86 & 6 & 2024 Nov 21 to 2024 Dec 18 & SPOC & \change{120} \\
\hline
\end{tabular}
\end{table*}

\subsection{TESS aperture contamination analysis}

\change{Flux from nearby stars can contaminate the target brightness (i.e., dilution). Due to the large pixel scale of TESS ($\sim$21\arcsec), the diluted transit depth has been shown to significantly affect TESS photometry leading to an underestimation and overestimation of $\sim6\%$ and $\sim20\%$ of the planet radius and density, respectively \citep{Han2025}. \changeblue{To evaluate the level of contamination in TOI-2155, we examined the TESS target pixel file \citep{2020A&A...635A.128A} together with contamination estimates from \texttt{tess-cont} \citep{2024A&A...691A.233C}.} }

\begin{figure*}
    \centering
     \begin{subfigure}{\columnwidth}
        \centering
        \includegraphics[width=\linewidth]{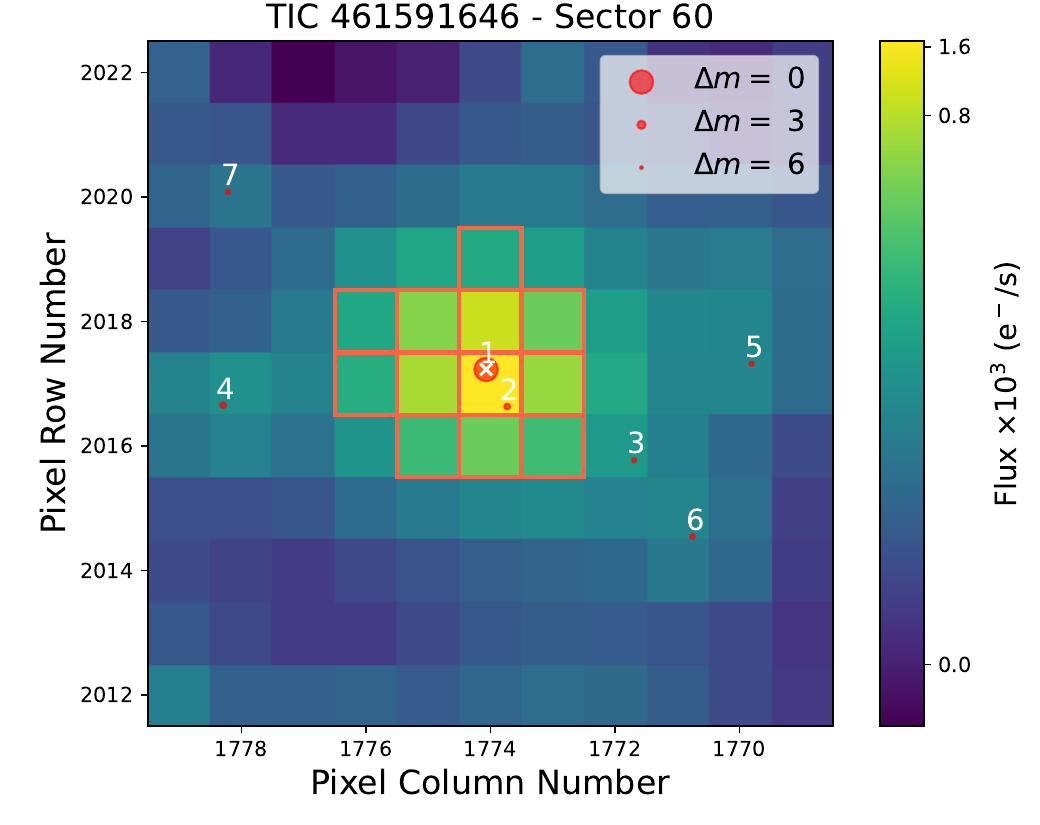}
      \caption{\changeblue{TESS target pixel file (TPF) of TOI-2155 with the SPOC aperture and nearby \textit{Gaia} DR3 sources overlaid.}}
        \label{fig:heatmap}
    \end{subfigure}
    \hfill
    \begin{subfigure}{\columnwidth}
        \centering
        \includegraphics[width=\linewidth]{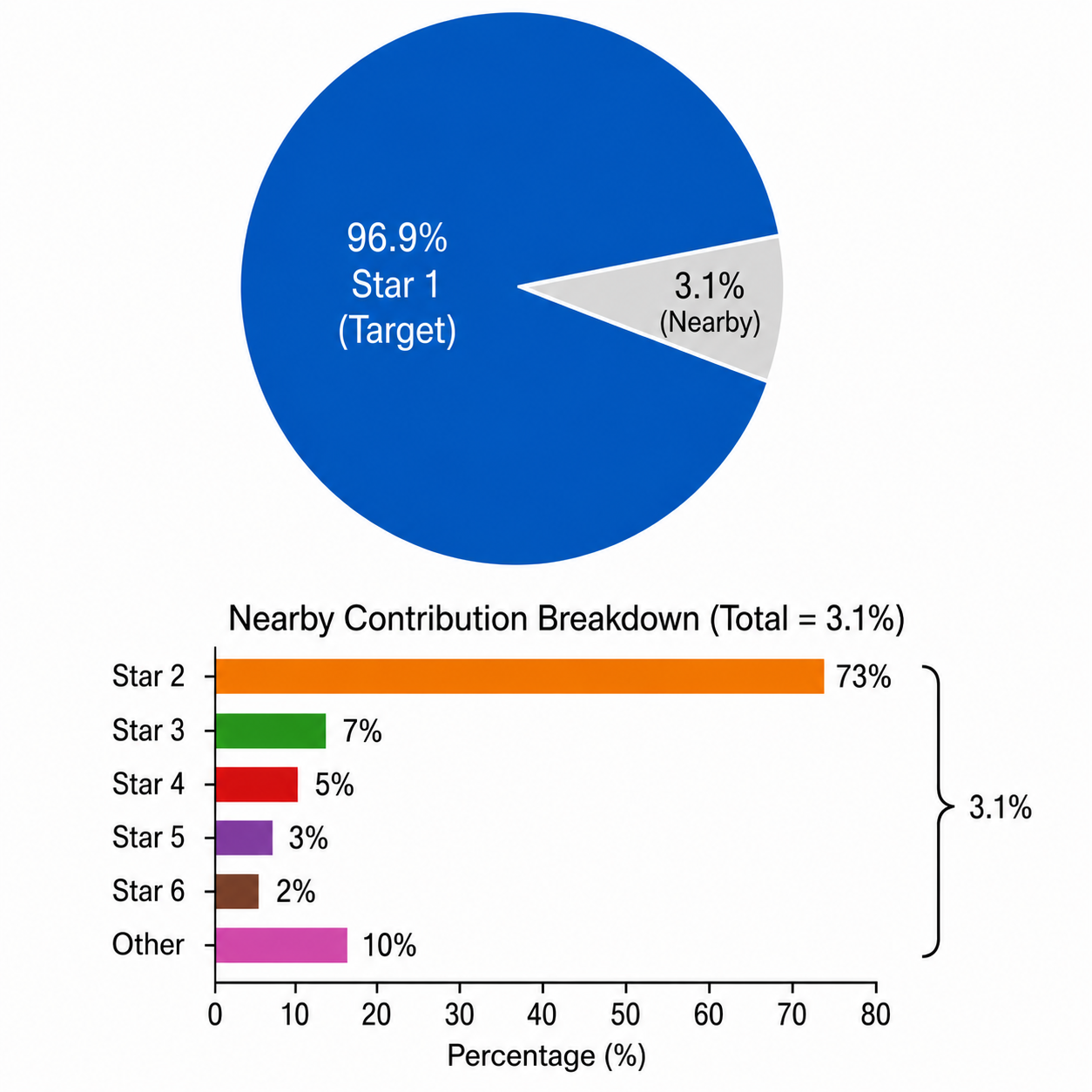}
      \caption{\changeblue{Flux contamination analysis of the TOI-2155 aperture from \texttt{tess-cont}.}}
        \label{fig:cont_panel}
    \end{subfigure}
     \caption{
   \changeblue{
(\textit{a}) TESS target pixel file (TPF) of TOI-2155 (TIC~461591646) from Sector~60, illustrating the pixel-level flux distribution within the field. 
The red outlines mark the SPOC photometric aperture, and the numbered markers identify nearby \textit{Gaia} DR3 sources, with marker size scaled by the relative brightness difference ($\Delta m$) with respect to the target. 
The target star (TOI-2155) lies near the center of the aperture, and the flux is strongly concentrated in the central pixels, indicating that the extracted photometry is dominated by the target star.
(\textit{b}) Flux contamination analysis from \texttt{tess-cont}, computed using the TESS pixel response function together with nearby \textit{Gaia} DR3 sources. 
The target star (TOI-2155) contributes $\sim96.9\%$ of the total flux within the adopted aperture, whereas nearby stars account for only $\sim3.1\%$ contamination. Among the nearby contaminating sources, Star~2 contributes the most flux to the aperture contamination.
}}
      \label{fig:tess_contamination}
\end{figure*}

In Figure~\ref{fig:heatmap}, we show the Sector~60 TESS target pixel file (TPF), with the SPOC photometric aperture overlaid on the median pixel flux distribution along with the locations of nearby \textit{Gaia} DR3 sources.

\change{The target star lies near the center of the aperture, and the pixel-level flux contribution indicates that it dominates the flux in the central pixels. Several neighboring Gaia sources are present within or near the aperture; however, all are significantly fainter than the target, and no comparably bright star within the aperture could account for the observed signal.}

\change{We further assessed the level of contamination using the crowding metrics provided by the SPOC pipeline. The crowding parameter CROWDSAP indicates that approximately 96--98\% of the flux within the aperture, across all available sectors, originates from the target star, with only a small contribution from nearby sources. The fraction of the target flux captured by the aperture (FLFRCSAP) is about 0.82--0.90, consistent with the expected TESS point-spread function.}

\changeblue{Overall, the level of flux contamination appears to be small. While the SPOC crowding metrics (CROWDSAP and FLFRCSAP) provide a useful indication, they may not capture all sources of contamination. To assess this independently, we used \texttt{tess-cont} \citep{2024A&A...691A.233C}, which models the flux contributions from nearby \textit{Gaia} DR3 sources using the TESS pixel response function. Figure~\ref{fig:cont_panel} presents the corresponding \texttt{tess-cont} contamination analysis for the SPOC aperture. The target contributes $\sim96.9\%$ of the flux within the aperture, corresponding to a contamination level of $\sim3.1\%$. This is consistent with the SPOC crowding metric (CROWDSAP $\approx$ 0.96--0.98), indicating that contamination remains at the few-percent level and is unlikely to significantly affect the derived system parameters.}

\subsection{Ground-based follow-up photometry}
\changeblue{Ground-based follow-up observations are used to confirm that the transit signal originates from the target star and to rule out nearby eclipsing systems or unresolved contaminants that could mimic or bias the signal. We confirm the transit through ground-based photometric follow-up observations. We further use high-resolution imaging as part of the follow-up to search for close companions at separations unresolved by TESS and \textit{Gaia}, providing an independent check for contaminating sources that could dilute the transit depth and bias the inferred radius of the transiting companion (see Section~\ref{high_resolution}).}

\subsubsection{Transit confirmation}
We observed a full transit on UTC 2021 May 31 in the Sloan $i'$ band from the Wendelstein Observatory (WST) in Munich, Germany. The 0.4\,m telescope is equipped with a QHY 600M camera with an image scale of $0\farcs264$, resulting in a $42\arcmin\times28\arcmin$ field of view. We used circular photometric apertures with a radius of $4\farcs5$.

We observed a full transit window on UTC 2021 September 13 in alternating Johnson/Cousins $B$ and Pan-STARRS $z_s$ bands from the Las Cumbres Observatory Global Telescope (LCOGT) \citep{Brown:2013} 1\,m network node at McDonald Observatory near Fort Davis, Texas, United States (McD). The 1\,m telescope is equipped with a $4096\times4096$ SINISTRO camera having an image scale of $0\farcs389$ per pixel, resulting in a $26\arcmin\times26\arcmin$ field of view. All images were calibrated by the standard LCOGT {\tt BANZAI} pipeline \citep{McCully:2018}. We used circular photometric apertures with a radius of $7\farcs0$ and $6\farcs2$, respectively.

\begin{figure}
    \centering
    \includegraphics[scale=0.40]{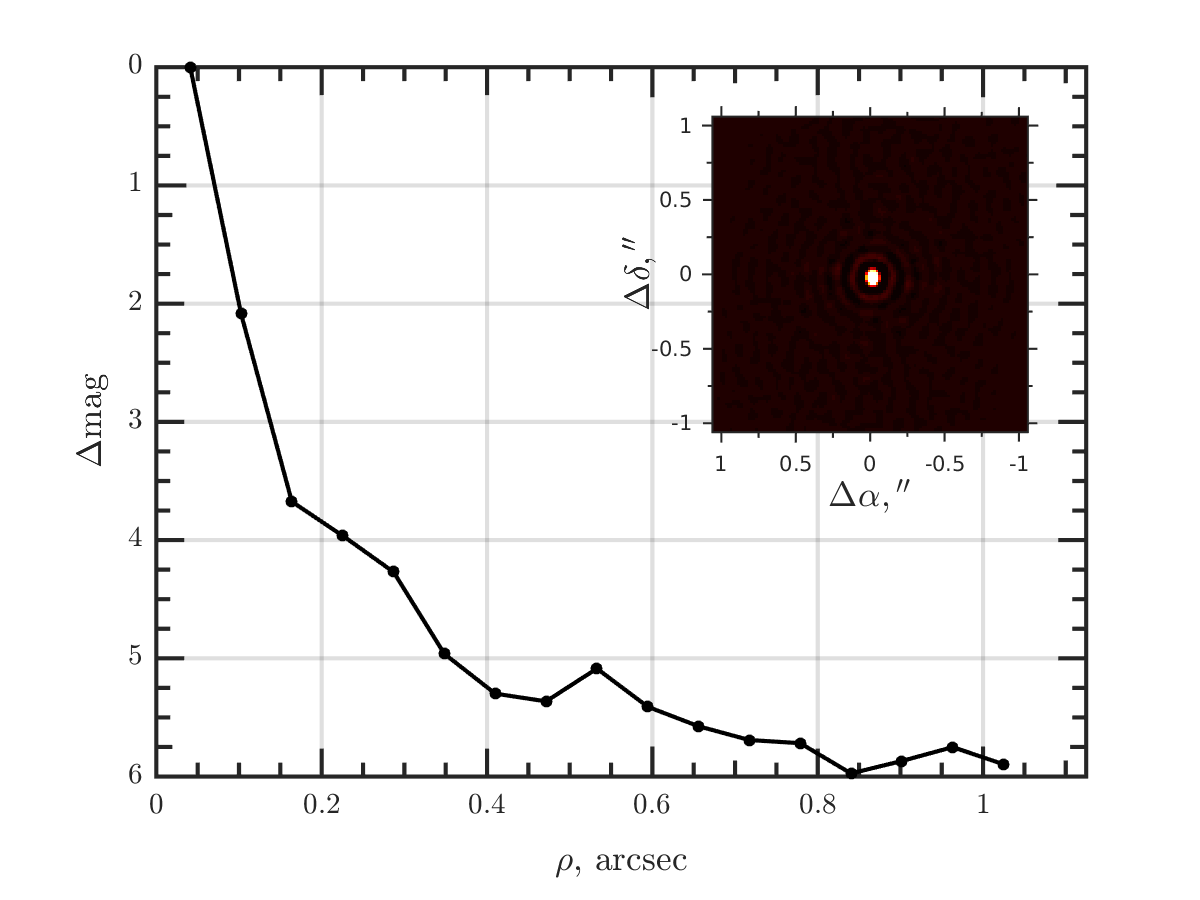}
    \caption{The 5$\sigma$ sensitivity limits of the SAI speckle observations of TOI-2155. The autocorrelation function is given in the inset. No nearby contaminating sources are detected.}
    \label{fig:sensitivity_curve}
\end{figure}

We observed one full transit window on UTC 2020 September 16 with no filter from the 0.24\,m telescope at Wild Boar Remote Observatory (WBR) in San Casciano in Val di Pesa, Firenze, Italy. The telescope is equipped with an SBIG ST8-XME detector with an image scale of 0$\farcs$79 pixel$^{-1}$, resulting in a $20\arcmin\times13.5\arcmin$ field of view. Image data from all observations were calibrated, and photometric data were extracted using {\tt AstroImageJ} \citep{Collins:2017}. We used circular photometric apertures with a radius of $7\farcs1$, which excluded all of the flux from the nearest known neighbor in the \textit{Gaia} DR3 catalog (Gaia DR3 573481114449495680) that is $14\farcs2$ northeast of TOI-2155. We scheduled our observations using the {\tt TESS Transit Finder}, a customized version of the {\tt Tapir} software package \citep{jensen2013tapir}.
An $\sim$4 ppt event was detected on-target in all four observations. All light curve data are available on the {\tt EXOFOP-TESS} website\footnote{\url{https://exofop.ipac.caltech.edu/tess/target.php?id=461591646}}.
\subsubsection{High-resolution imaging and contaminating sources}
\label{high_resolution}
\changeblue{To further constrain potential sources of contamination, we obtained high-resolution imaging to search for close companions at angular separations unresolved by TESS and \textit{Gaia}. These observations are conducted at angular resolutions beyond the seeing limit, allowing us to detect  close companions and assess whether they contaminate the photometry of the target star.}

We observed TOI-2155 on October 24, 2020, at 22:48 UT with the Speckle Polarimeter \citep{safonov2017speckle} on the 2.5~m telescope at the Caucasian Mountain Observatory (CMO) of the Sternberg Astronomical Institute (SAI) of Lomonosov Moscow State University. The CMO is located in the northern Caucasus, and the telescope coordinates are $43^{\circ}44^{\prime}10^{\prime\prime}  N, 42^{\circ}40^{\prime}03^{\prime\prime}$
E, 2100 m above sea level. An Electron-Multiplying CCD, the Andor iXon 897, was used as a detector. The atmospheric dispersion compensator allowed observation through the wide-band $I_c$ filter. The power spectrum was estimated using 4000 frames with a 30-ms exposure. The detector has a pixel scale of $20.6$ mas pixel$^{-1}$, and the angular resolution was 89 mas. The long-exposure vision was $1.45^{\farcs}$. We did not detect any stellar companions brighter than $\Delta I_c = 4.1$ and $5.9$ at angular separations of $\rho = 0\farcs25$ and $1\farcs0$, respectively, where $\rho$ is the separation between the source and the potential companion, \change{corresponding to physical separations of approximately 104~au and 414~au for TOI-2155}. Given TESS's large pixel scale of 21$\arcsec$, our high-resolution imaging, with an angular resolution of 89 milliarcseconds, effectively excludes nearby companions that could contaminate the photometry.

Using a custom pipeline to process data from the speckle polarimeter \citep{tokovinin2010speckle}, we show in Figure~\ref{fig:sensitivity_curve} the 5$\sigma$ sensitivity limits of the SAI speckle observations of TOI-2155, where no nearby contaminating sources are detected.  While we cannot constrain companions beyond 1.2~arcsec due to the limited field of view, \textit{Gaia} is sensitive to stellar companions at these larger separations and would detect any such objects.

\begin{deluxetable}{ccc}
\tablecaption{Radial Velocity measurements of TOI-2155 obtained with TRES (in m~s$^{-1}$). \label{table:rvdata}}
\tablehead{
\colhead{Time} & \colhead{RV} & \colhead{{$\sigma_{\rm RV}$}} \\
\colhead{(BJD)} & \colhead{(m s$^{-1}$)} & \colhead{(m s$^{-1}$)}
}
\startdata
2459113.834178 &  -9399.9 & 187.6 \\
2459115.946682 &   6627.2 &  86.7 \\
2459117.791870 & -10603.6 & 116.0 \\
2459126.791904 &   5867.1 &  82.1 \\
2459127.793411 &   2739.9 &  94.0 \\
2459128.777122 &  -9596.5 & 101.3 \\
2459130.854558 &   6994.7 & 102.7 \\
2459131.876366 &  -3074.3 &  75.8 \\
2459132.790768 & -10548.3 &  82.4 \\
2459133.785629 &    -57.7 &  76.7 \\
2459861.726508 &      1.3 &  66.1 \\
2459861.742700 &      8.9 &  73.5 \\
2459861.762933 &   -511.3 &  73.9 \\
2459861.779137 &   -691.1 &  62.1 \\
2459861.795463 &   -933.7 &  63.2 \\
2459861.812957 &  -1090.5 &  61.0 \\
2459861.829619 &  -1504.2 &  67.8 \\
2459861.845788 &  -1648.2 &  93.8 \\
2459861.862097 &  -1971.5 &  83.4 \\
2459861.873585 &  -1990.6 & 156.6 \\
2459861.885935 &  -2163.4 &  81.2 \\
2459861.902498 &  -2039.7 & 150.5 \\
2459861.919124 &  -2327.4 & 111.1 \\
2459861.937267 &  -2787.4 & 140.3 \\
2459861.953807 &  -3015.0 &  93.0 \\
2459861.970480 &  -3148.4 &  98.4 \\
2459861.987281 &  -3658.9 & 113.3 \\
\enddata
\end{deluxetable}


\subsubsection{Radial velocities}

To characterize the TOI-2155 system and confirm the nature of its transiting companion, we obtained a total of 27 spectra using the Tillinghast Reflector Echelle Spectrograph (TRES) \citep{gaborthesis} on the 1.5-meter telescope at the Fred L. Whipple Observatory on Mount Hopkins, Arizona. These 27 spectra were used to measure the radial velocity (RV) variations induced by the companion, allowing us to determine its mass. 
TRES  covers a wavelength range of 3850 to 9096~\AA\ and has a resolving power of approximately $R \sim 44{,}000$. The TRES team measured relative RVs at each orbital phase using multiple \'{e}chelle orders per spectrum \citep{buchhave2010}. Cosmic rays were identified and removed through visual inspection of each order, and low signal-to-noise orders were excluded, along with those contaminated by telluric absorption. Each spectrum's radial velocity (RV) was determined by cross-correlating the individual \'{e}chelle orders against a median co-added template constructed from the out-of-transit spectra, with outlier pixels replaced during the template creation. The relative RV for each observation was computed as the average of the velocities from all usable orders. Finally, the measurements were shifted to correct for long-term drift in the TRES zero point, which is monitored nightly via RV standard stars \citep[similar to the process outlined in][]{quinn2014}. The RV uncertainties are calculated from the standard deviations of the RVs of the individual \'{e}chelle orders, and inflated by the instrumental jitter measured in the scatter of the standard star RVs (typically 10 to 15 m s$^{-1}$).
\change{We obtained a series of radial-velocity measurements of TOI-2155 with the TRES spectrograph, spanning a time baseline of approximately two years from September 2020 to October 2022. Two early spectra were obtained on September 23 and October 11, 2020, with exposure times of 1300–2400 s and signal-to-noise ratios of 27.6-31.9. Additional TRES observations were subsequently acquired to sample the orbit over a longer time interval and refine the radial-velocity solution}. The complete set of RVs derived from TRES is listed in Table~\ref{table:rvdata}.

\section{Analysis and Results}
\label{sec: results}

\subsection{Stellar properties}

The effective temperature $T_{\rm eff}$, metallicity [Fe/H], surface gravity $\log g$, and projected stellar rotational velocity $v \sin I_{\star}$ were derived by applying the Stellar Parameter Classification (SPC) software \citep{bieryla2021stellar,buchhave2012} to each TRES spectrum individually. The final parameters were obtained by averaging the results from all spectra. The SPC uses a library of computed spectra in the wavelength region of $5020 \, \text{\AA} \text{ to } 5320 \, \text{\AA}$, centered on the Mg b triplet. 

We obtained the following stellar parameters for TOI-2155 using SPC: $T_{\text {eff }}=$ $6085 \pm 78 \mathrm{~K}, \log g=4.21 \pm 0.13,[\mathrm{M} / \mathrm{H}]=0.13\pm 0.08$, and $v \sin I_{\star}=19.8\pm 0.5\mathrm{~km} \mathrm{~s}^{-1}$ all of which are summarised in Table~\ref{tab:toi2155_stellar_params}. This Table also lists the proper motion, parallax, and magnitudes from Gaia DR3 \citep{collaboration2023gaia},
TESS, 2MASS \citep{skrutskie2006two}, and SPC \citep{bieryla2021stellar}.
Using a bolometric correction of \( \mathrm{BC}_G = 0.0 \), the bolometric magnitude is \( M_{\mathrm{bol}} = M_G + \mathrm{BC}_G \). The bolometric luminosity was then calculated using
\begin{equation}
\log_{10}\left(\frac{L}{L_\odot}\right) = -0.4 \left(M_G + \mathrm{BC}_G - M_{\mathrm{bol},\odot}\right),
\end{equation}
where $M_{\mathrm{bol},\odot} = 4.74$ is the bolometric magnitude of the Sun. Propagating the uncertainties in distance and apparent magnitude, we find a luminosity of $3.74 \pm 0.04$ ${L_\odot}$, consistent with an F-type dwarf \citep{Pecaut_2013}.

\begin{table*}
\centering
\caption{Astrometry, photometry, spectroscopy, and derived stellar properties for TOI-2155. Sources: \textbf{(1)} Gaia EDR3 \citep{gaia2021}, \textbf{(2)} \textit{TESS} Input Catalog \citep{2018AJ....156..102S}, \textbf{(3)} 2MASS \citep{skrutskie2006two}, \textbf{(4)} WISE \citep{WISE}, \textbf{(5)} Tycho \citep{tycho2}, \textbf{(6)} GALEX \citep{bianchi2012vizier}}.
\begin{tabular}{|l|p{3.5cm}|p{2cm}|}
\hline
\multicolumn{3}{|l|}{\textbf{Star Information}}\\
\hline
\multicolumn{3}{|l|}{\textbf{Target Designations}} \\
\hline
TOI ID & 2155 & \\
TIC ID & 461591646 & \\
Gaia DR3 ID & \texttt{573481114449301376} & \\
2MASS ID & \texttt{J00231182+8427102} & \\
\hline
\textbf{Parameter} & \textbf{Value} & \textbf{Source} \\
\hline

\multicolumn{3}{|l|}{\textbf{Astrometric Properties}} \\
\hline
Right Ascension (RA) & 00:23:12.06 & (1) \\
Declination (Dec) & +84:27:10.22 & (1) \\
Proper Motion in RA ($\mu_{\alpha}$) [mas yr$^{-1}$] & $22.3 \pm 0.01$ & (1) \\
Proper Motion in Dec ($\mu_{\delta}$) [mas yr$^{-1}$] & $-1.5 \pm 0.02$ & (1) \\
Parallax [mas] & $2.416 \pm 0.012$ & (1) \\
Distance [pc] & $414 \pm 2.0$ & (1) \\
\hline

\multicolumn{3}{|l|}{\textbf{Photometric Properties}} \\
\hline
Gaia $G$ [mag] & $11.395 \pm 0.020$ & (1) \\
Gaia $G_{\rm BP}$ [mag] & $10.942 \pm 0.020$ & (1) \\
Gaia $G_{\rm RP}$ [mag] & $11.705 \pm 0.020$ & (1) \\
TESS [mag] & $10.994 \pm 0.010$ & (2) \\
2MASS $J$ [mag] & $10.42 \pm 0.03$ & (3) \\
2MASS $H$ [mag] & $10.16 \pm 0.03$ & (3) \\
2MASS $K_S$ [mag] & $10.10 \pm 0.02$ & (3) \\
WISE1 (3.4 $\mu$m) [mag] & $10.082 \pm 0.030$ & (4) \\
WISE2 (4.6 $\mu$m) [mag] & $10.105 \pm 0.030$ & (4) \\
WISE3 (12 $\mu$m) [mag] & $10.088 \pm 0.048$ & (4) \\
$B_T$ [mag] & $11.920 \pm 0.091$ & (5) \\
$V_T$ [mag] & $11.748 \pm 0.130$ & (5) \\
\change{GALEX NUV [mag]} &  \change{15.968} $\pm$ \change{0.013} & \change{(6)} \\
\hline

\multicolumn{3}{|l|}{\textbf{Spectroscopic and Derived Parameters}} \\
\hline
\change{Spectral type} & \change{F8--G0\,IV/V} & \change{this work} \\
Effective Temperature ($T_{\mathrm{eff}}$) [K] & $6085 \pm 78$ & this work \\
Surface Gravity ($\log g$) [dex] & $4.21 \pm 0.13$ & this work \\
Metallicity ([Fe/H]) [dex] & $0.13 \pm 0.08$ & this work \\
\change{Projected rotational velocity ($v \sin i$) [km s$^{-1}$]} & \change{$19.8\pm 0.5$} & \change{this work }\\
\change{Projected rotation period ($P_{\rm rot}/\sin i$) [days]} & \change{$4.4 \pm 0.1$} & \change{this work} \\
Stellar Mass ($M_{\odot}$) & $1.33 \pm 0.008$ & this work \\
Stellar Radius ($R_{\odot}$) & $1.705^{+0.066}_{-0.064}$ & this work \\
Luminosity ($L_{\odot}$) & $3.74 \pm 0.04$ & this work \\
Stellar Density [g cm$^{-3}$] & $0.390^{+0.033}_{-0.030}$ & this work \\
Age [Gyr] & $3.2^{+1.9}_{-0.9}$ & this work \\
\hline

\end{tabular}
\label{tab:toi2155_stellar_params}
\end{table*}
\subsubsection{Stellar mass, radius, and age}

\begin{figure}
    \centering
    \includegraphics[width=\columnwidth]{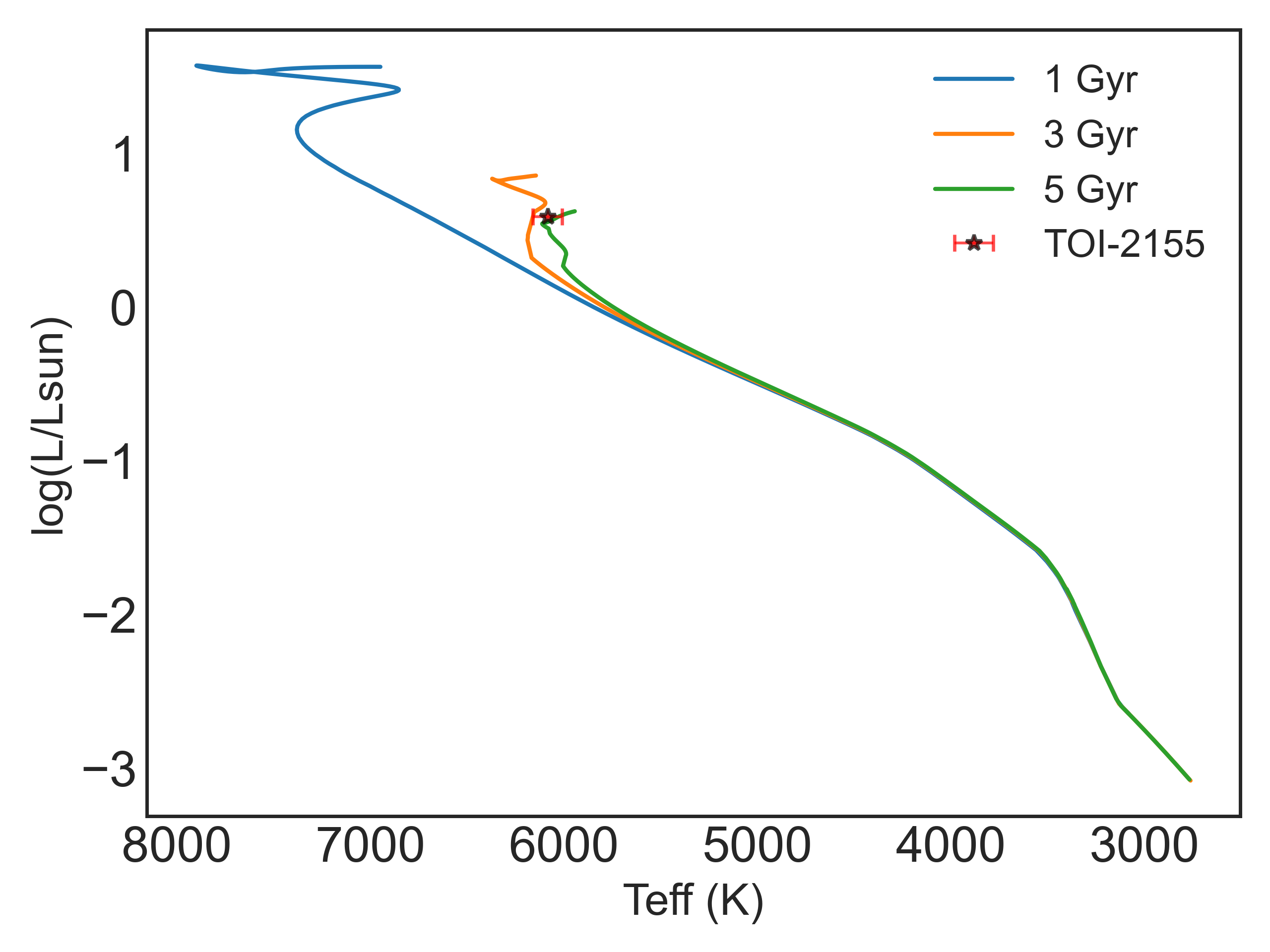}
 \caption{\changeblue{
Position of TOI-2155 in the $T_{\rm eff}$--luminosity plane. 
The luminosity is derived from SED modelling. 
MIST isochrones at 1, 3, and 5 Gyr (using the nearest available grid values) \changegreen{with metallicity $[{\rm Fe/H}] = 0.1$} are overlaid to illustrate the evolutionary state of the star. 
The location of TOI-2155 lies between the 3 and 5 Gyr isochrones, consistent with the inferred age of $3.2^{+1.9}_{-0.9}$ Gyr. 
The horizontal error bar represents the uncertainty in effective temperature.
}}
    \label{fig:mist_isochrone}
\end{figure}

\change{We estimated the stellar properties of TOI-2155 using two complementary approaches: 
(1) a global fit with MESA Isochrones and Stellar Tracks (MIST) models \citep{mist1, mist2, mist3} implemented in {\tt EXOFASTv2} \citep{eastman2019}, and 
(2) an independent empirical spectral energy distribution (SED) analysis. We adopt the EXOFASTv2 solution as the primary result, with the empirical SED analysis used as an independent validation}.
{\tt EXOFASTv2} uses a Differential Evolution Markov Chain Monte Carlo (MCMC) method to jointly fit a suite of parameters to a star or exoplanet system. To create a spectral energy distribution (SED) model for the star and to assess its age, we establish a set of priors for the MCMC fit. We used our spectroscopic measurements of [Fe/H] and $T_{\rm eff}$ and parallax measurements from \textit{Gaia} (with the zero-point offset accounted for \citep{dr3_correction} to define Gaussian priors. We impose an upper limit with uniform priors for the $A_V$ extinction taken from \cite{Schlegel:1998}. We use the \textit{Gaia} $G$, \textit{Gaia} $Bp$, \textit{Gaia} $Rp$, $B$, $V$, $J$, $H$, $Ks$, and WISE $W1$, $W2$, and $W3$ photometry bands to construct the SED model. \change{The TESS magnitude was not included in the SED fits because it was not required, given the available broadband photometry}. The SED photometry values are given in Table \ref{tab:toi2155_stellar_params}. Using {\tt EXOFASTv2}, we estimate the Age $= 3.2^{+1.9}_{-0.93}$\,Gyr and the radius of the host star $R_\star =1.705^{+0.066}_{-0.064}\,\mathrm{R_\odot}$. These {\tt EXOFASTv2} values are adopted throughout this work. \changeblue{Figure~\ref{fig:mist_isochrone} shows the location of TOI-2155 in the $T_{\rm eff}$--luminosity plane, together with MIST isochrones at 1, 3, and 5 Gyr. The star lies between the 3 and 5 Gyr isochrones, consistent with the inferred age of $3.2^{+1.9}_{-0.9}$ Gyr.}

\begin{table*}
\centering
\caption{Fitted parameters for TOI-2155\,b from \texttt{Allesfitter}. Priors are shown as uniform $\mathcal{U}$. The values are the median from the posterior distribution. }
\begin{tabular}{llll}
\toprule
Parameter & Symbol & Prior, parameter space & Value \\
\hline
\multicolumn{4}{l}{\textbf{System parameters}} \\
Radius ratio & $R_b / R_{\star}$ & $\mathcal{U}(0, 0.1)$ & \change{$0.0586^{+0.00048}_{-0.00057}$} \\
Scaled summed radius & $(R_{\star}+R_b)/a_b$ & $\mathcal{U}(0, 0.3)$ & \change{$0.164^{+0.0045}_{-0.0048}$} \\
Cosine inclination & $\cos i_b$ & $\mathcal{U}(0, 0.3)$ & \change{$0.111\pm0.0065$} \\
Epoch (BJD) & $T_{0;b}$ & $\mathcal{U}(2459891, 2459892)$ & \change{$2459724.02337\pm0.00024$}  \\
Orbital period (d) & $P_b$ & $\mathcal{U}(3.65, 3.80)$ & \change{$3.7246950 \pm 0.0000014$} \\
RV Semi-amplitude (km\,s$^{-1}$) & $K_b$ & $\mathcal{U}(8, 9)$ & \change{$8.74\pm0.12$} \\
$\sqrt{e_b}\cos\omega_b$ & $f_c$ & $\mathcal{U}(-1, 1)$ & \change{$0.059_{-0.061}^{+0.047}$} \\
$\sqrt{e_b}\sin\omega_b$ & $f_s$ & $\mathcal{U}(-1, 1)$ & \change{$-0.06_{-0.08}^{+0.10}$} \\

\multicolumn{4}{l}{\textbf{Limb darkening parameters}} \\
TESS & $q_{1;\mathrm{TESS}}$ & $\mathcal{U}(0, 1)$ & \change{$0.17_{-0.05}^{+0.06}$}  \\
TESS & $q_{2;\mathrm{TESS}}$ & $\mathcal{U}(0, 1)$ & \change{$0.30_{-0.20}^{+0.32}$} \\

\multicolumn{4}{l}{\textbf{White noise parameters}} \\
TESS white noise & $\log \sigma_{\mathrm{TESS}}$ & $\mathcal{U}(-8, -4)$ & \change{$-6.361\pm0.0064$}  \\
RV jitter (TRES) & $\ln \sigma_{\text{jitter}}(RV_{\mathrm{TRES}})$ & $\mathcal{U}(-2, 0)$ & \change{$-1.27_{-0.18}^{+0.20}$}  \\

\multicolumn{4}{l}{\textbf{Baseline parameters}} \\
TESS offset & $\Delta_{\mathrm{TESS}}$ & $\mathcal{U}(-0.5, 0.5)$ &  \change{$0.000581\pm0.000018$} \\
TRES offset & $\Delta_{\mathrm{TRES}}$ & $\mathcal{U}(-2, 2)$ &  \change{$-1.60_{-0.09}^{+0.09}$} \\
\hline
\bottomrule
\end{tabular}
\label{fitted_params}
\end{table*}

\begin{figure}
    \includegraphics[scale=0.30]{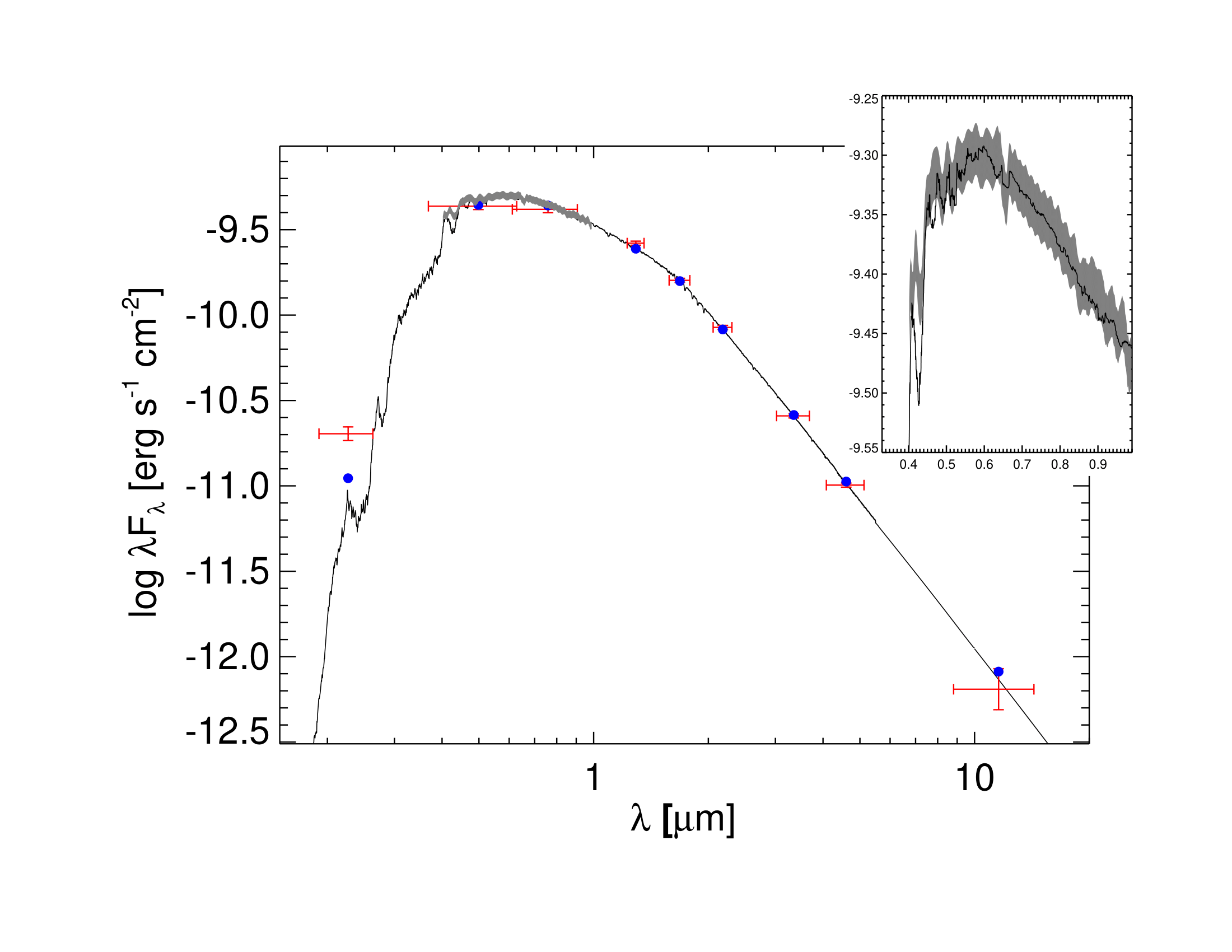}
    \caption{Spectral energy distribution of TOI-2155. Red symbols represent the observed photometric measurements, with horizontal bars indicating the effective passband width. Blue symbols are the model fluxes from the best-fit PHOENIX atmosphere model (black). The inset shows the {\it \textit{Gaia}\/} spectrophotometry overlaid as a grey swathe.}
    \label{phoneix_sed}
\end{figure}

As an independent determination of the basic stellar parameters, we performed an analysis of the broadband SED of the star together with the {\it \textit{Gaia}\/} DR3 parallax without systematic offset applied \citep[see e.g.][]{stassun2021parallax}, to determine an empirical measurement of the stellar radius, following the procedures described in \citet{stassun2016eclipsing,stassun2017empirical,stassun2018evidence}. We extracted the $JHK_S$ magnitudes from {\it 2MASS}, the W1--W3 magnitudes from {\it WISE}, the $G_{\rm BP} G_{\rm RP}$ magnitudes from {\it \textit{Gaia}}, and the NUV magnitude from {\it GALEX}. We also used absolute flux-calibrated {\it \textit{Gaia}\/} spectrophotometry. Together, the available photometry spans the entire stellar SED over the wavelength range 0.2--10~$\mu$m (see Figure~\ref{phoneix_sed}).   
 
We performed a fit using PHOENIX stellar atmosphere models \citep{husser2013new}, with $T_{\rm eff}$, $\log g$, and [Fe/H] adopted from the spectroscopic analysis. The extinction $A_V$ was limited to the maximum line-of-sight value from the Galactic dust maps of \citet{Schlegel:1998}. The resulting fit (Figure~\ref{phoneix_sed}) has a best-fit $A_V = 0.16 \pm 0.08$, with a reduced $\chi^2$ of 1.4. Integrating the (unreddened) model SED gives the bolometric flux at Earth, $F_{\rm bol} = 7.27 \pm 0.26 \times 10^{-10}$ erg~s$^{-1}$~cm$^{-2}$. Taking the $F_{\rm bol}$ together with the {\it \textit{Gaia}\/} parallax directly gives the bolometric luminosity, $L_{\rm bol} = 3.89 \pm 0.14$~L$_\odot$. 
\change{The stellar radius follows from the Stefan-Boltzmann relation, giving $R_\star = 1.76 \pm 0.03$~R$_\odot$}.
In addition, we can estimate the stellar mass from the empirical relations of \citet{torres2010use}, giving $M_\star = 1.33 \pm 0.008$~M$_\odot$.
\change{Finally, we estimate the (projected) rotation period from the spectroscopic $v\sin i$ together with $R_\star$ from above, giving $P_{\rm rot}/\sin i = 4.4 \pm 0.1$~days}. \changeblue{We also find no detectable lithium absorption at 6708\,\AA, consistent with the inferred mature age of the system}.

\begin{figure}
    \centering
    \includegraphics[scale=0.40]{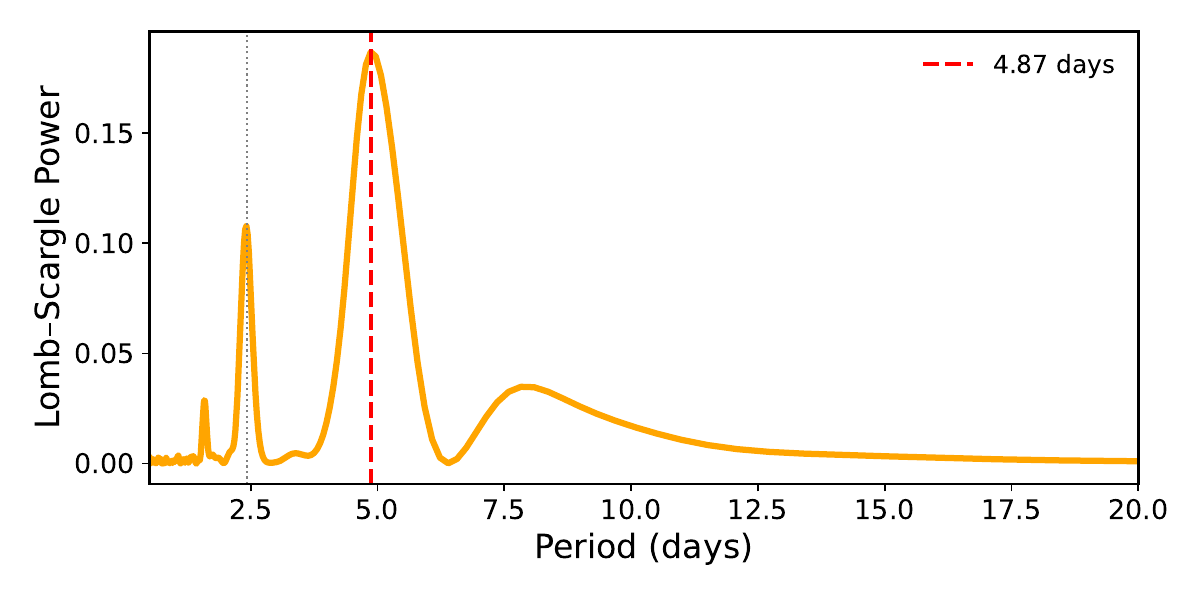}
    \caption{Lomb--Scargle periodogram of the out-of-transit TESS light curve of TOI-2155 of sector~52. 
The dominant peak occurs near $4.87$\, days and may correspond to the stellar rotation period. 
Signals near half this period could represent harmonic variability.}
    \label{fig:lc_periodogram}
\end{figure}

\change{The stellar radius derived from the Stefan–Boltzmann relation is consistent within uncertainties with the value obtained from {\tt EXOFASTv2}, supporting the robustness of the adopted stellar parameters}.

\begin{figure*}
    \centering
    \begin{subfigure}[t]{0.49\textwidth}
        \centering
        \includegraphics[height=8cm]{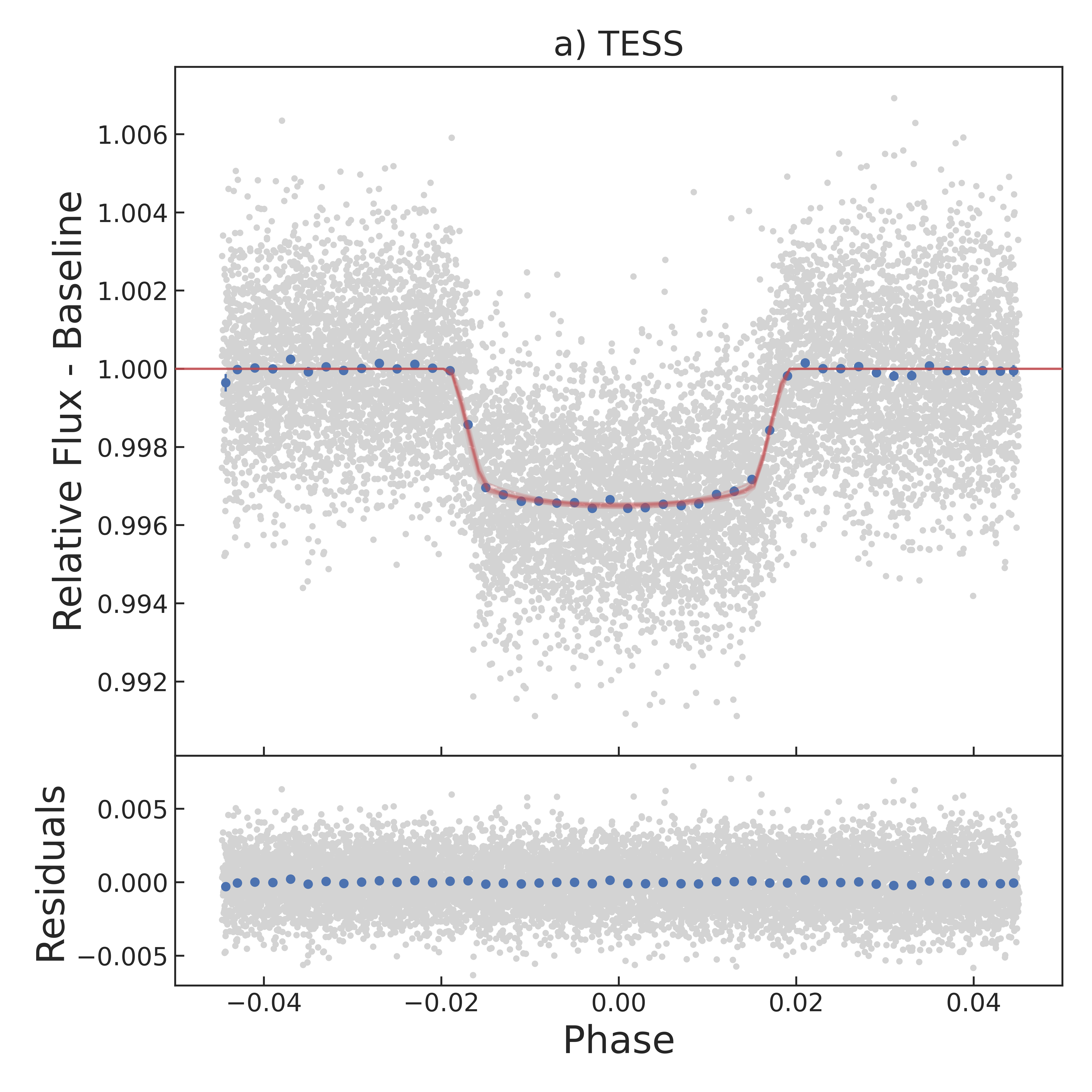}
        \vspace{2pt}
    \end{subfigure}
    \hfill
    \begin{subfigure}[t]{0.48\textwidth}
        \centering
        \includegraphics[height=8cm]{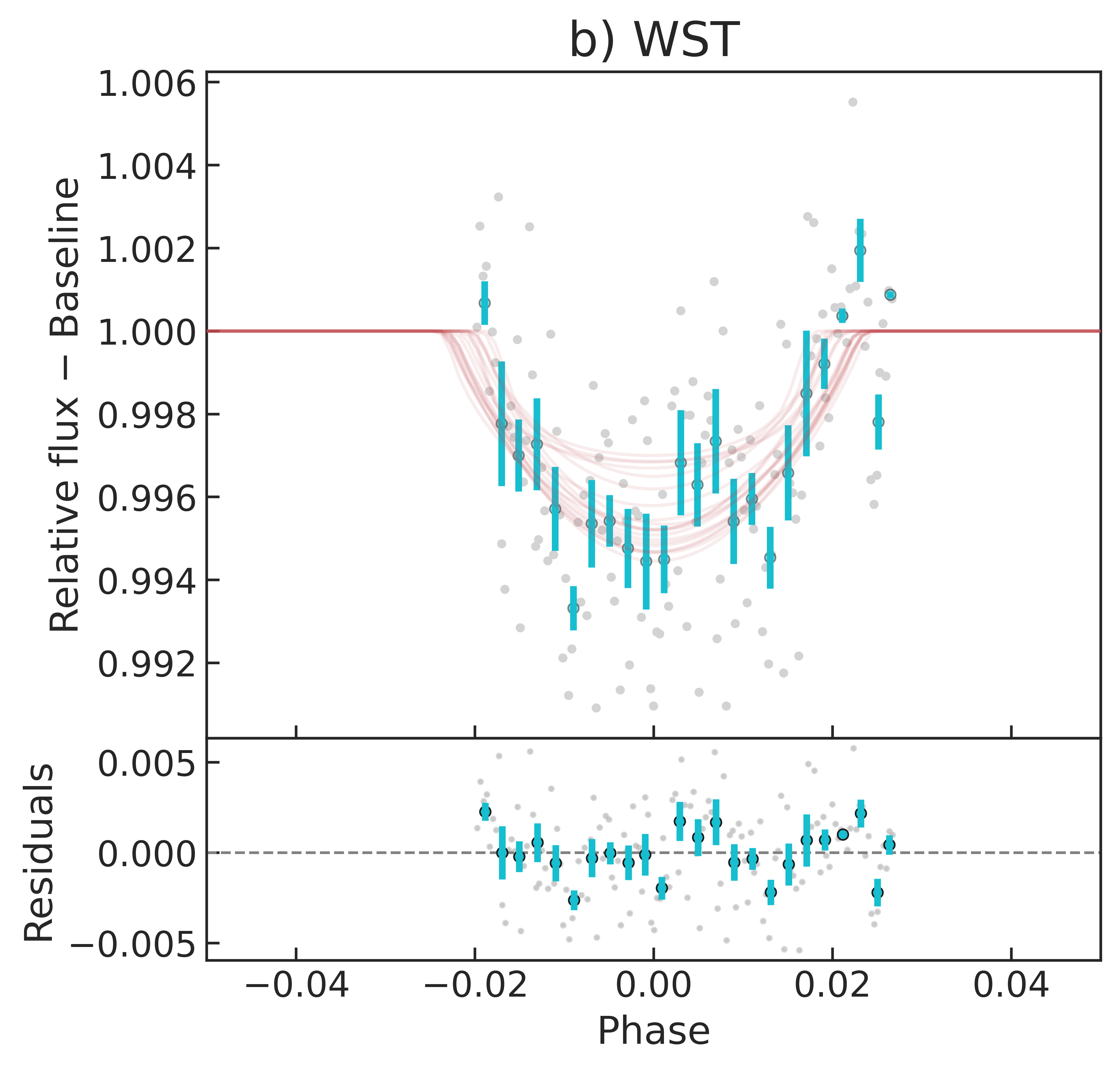}
        \vspace{2pt}
    \end{subfigure}

    \vspace{8pt} 

    \begin{subfigure}[t]{0.48\textwidth}
        \centering
        \includegraphics[height=8cm]{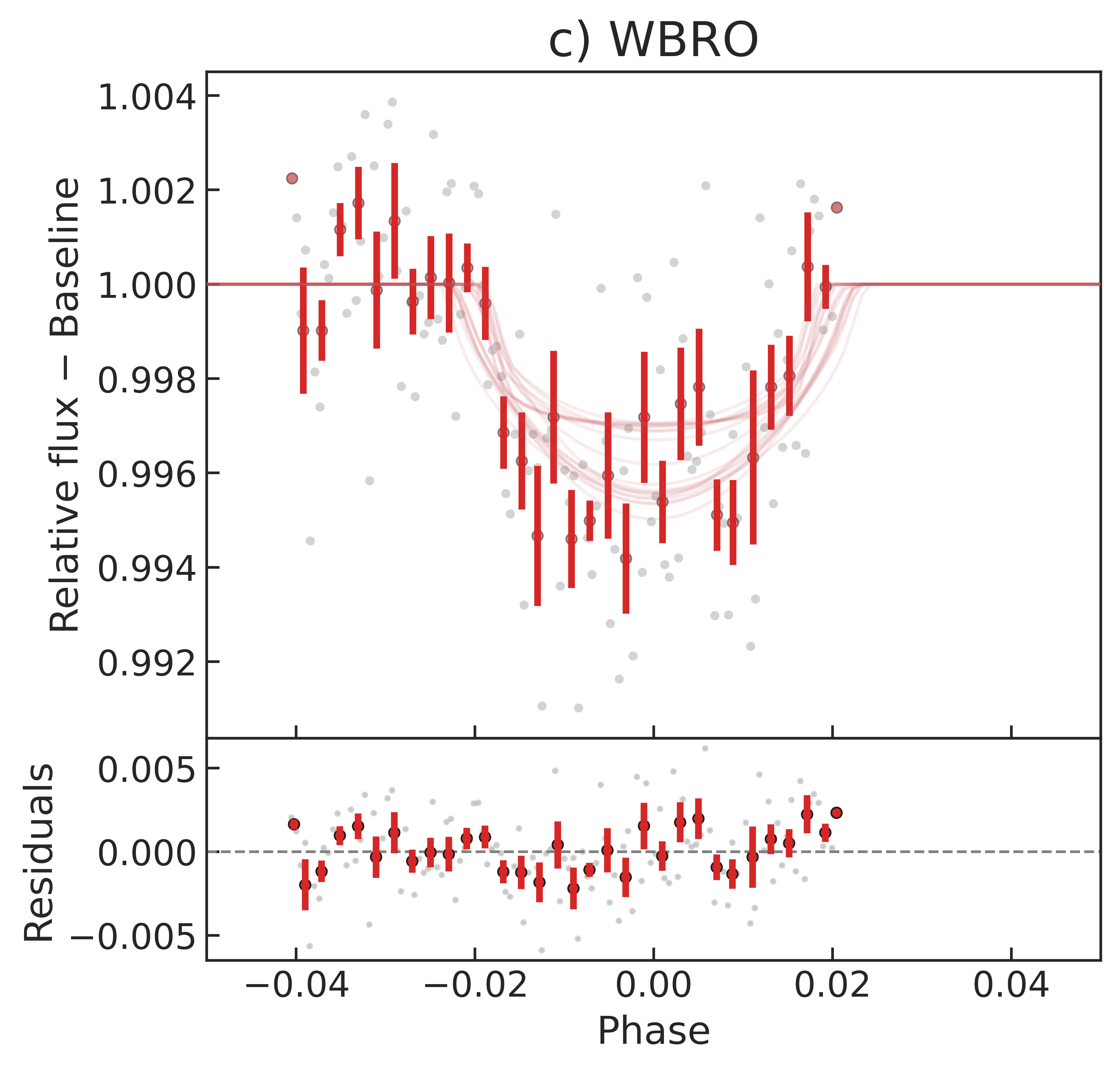}
        \vspace{2pt}
    \end{subfigure}
    \hfill
    \begin{subfigure}[t]{0.48\textwidth}
        \centering
        \includegraphics[height=8cm]{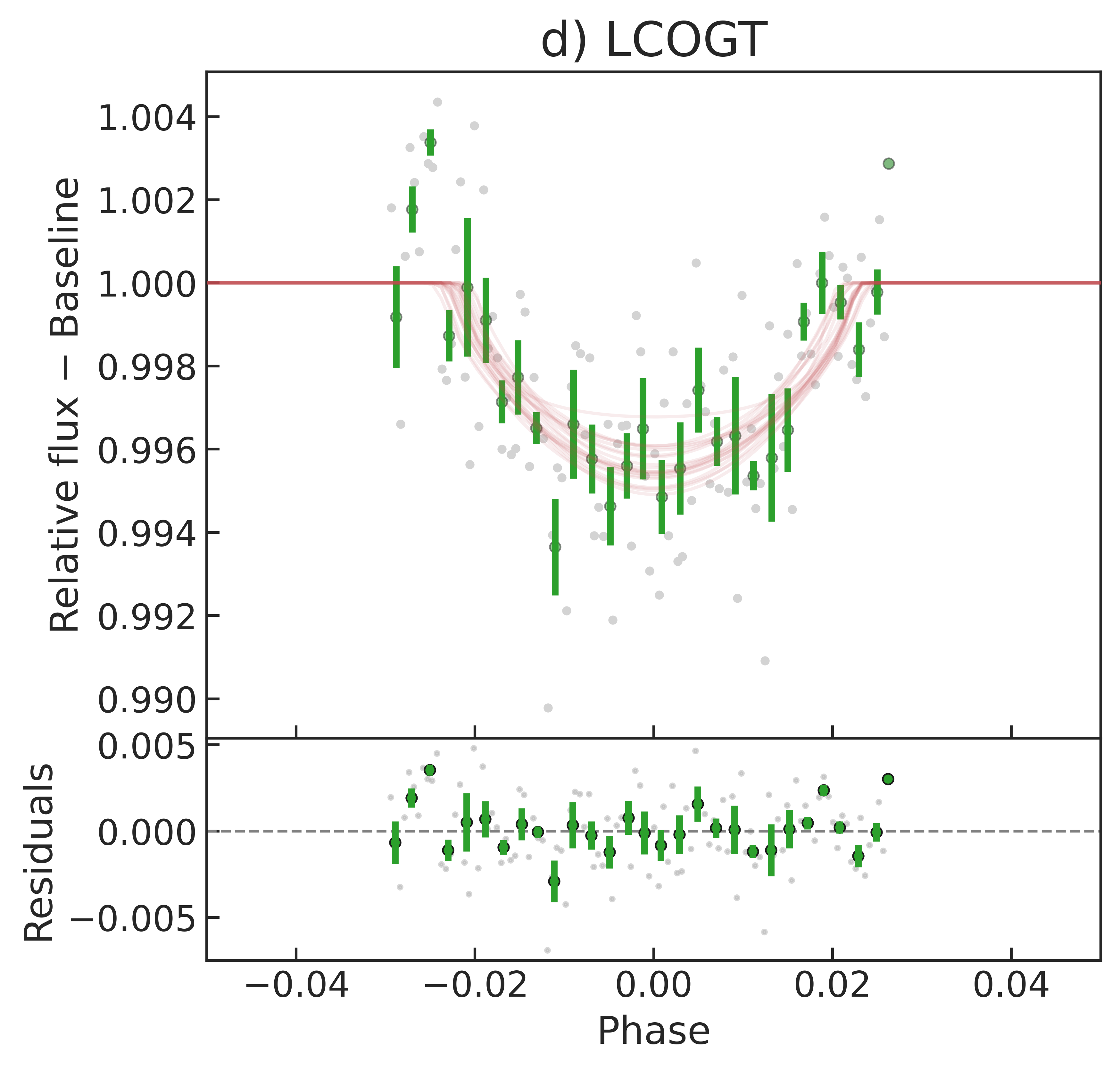}
        \vspace{2pt}
    \end{subfigure}

    \caption{%
        \textbf{(a)} TESS phase-folded data (gray points) with Allesfitter model (red curve), and binned data (blue points). The lower panel shows the residuals after subtracting the model and baseline from the relative flux. 
        \textbf{(b–d)} Phase-folded ground-based photometry data from WST, WBRO, and LCOGT (cyan, red, and green points), model attempt (red curve) from \texttt{Allesfitter}, and \change{the residuals in the lower panels follow the same colour scheme as the upper panels, clearly separating raw and binned data}. In the top panel of all plots (a–d), we remove the baseline from the relative flux, which represents an offset. The ground-based photometry data points are not used in global modeling as the data don't have a sufficient baseline \change{and do not consistently capture both ingress and egress (see Section~3.2)}.
    }
    \label{fig: all_phaseplots}
\end{figure*}

\subsubsection{Photometric modulation and stellar rotation period}


During initial analysis of TESS light curves, we found photometric modulation. \change{We investigated the out-of-transit data from all sectors and use the Lomb-Scargle periodogram  \citep{lomb1976least,scargle1982studies}. The analysis revealed a signal near $\sim$5\, days (Figure~\ref{fig:lc_periodogram}) that also appears in several individual sectors, including sectors 52, 59, 78, and 86. We further examined the continuous sectors 18--20, which show signals near $\sim$4.8\,days and $\sim$2.4--2.5\,days respectively. The shorter period may represent a harmonic of the longer one, which can occur when stellar surface features produce two brightness variations per rotation. Periods near $\sim$5\, days can also coincide with known TESS spacecraft systematics, such as the first harmonic of the $\sim$2.5-day momentum dump cadence \citep{10.1093/mnras/staa814}. However, the dominant signal at 4.87\,days is slightly offset from this exact harmonic and is consistently detected across multiple sectors, supporting an astrophysical origin. We note that the recent paper by \cite{boyle2026tess} reports a rotation period of $4.876 \pm 0.063$\, days for TIC~461591646, consistent with the signal identified in our analysis. We did not analyse data from the All-Sky Automated Survey for SuperNovae (ASAS-SN; \citealt{Holoien2017}), as its longer cadence and lower photometric precision compared to TESS would make the rotation signal difficult to detect.}


\changeblue{Combining the measured rotation period with the stellar radius and spectroscopic $v \sin i$, we can estimate the stellar inclination following \cite{masuda2020inference}. Using the parameters listed in Table~\ref{tab:toi2155_stellar_params}, we find $\sin i \approx 1$, indicating that the stellar rotation axis is nearly perpendicular to the line of sight, and that the stellar equatorial plane is viewed close to edge-on.}

\changeblue{We also consider the stellar rotation in the context of gyrochronology. For stars near the F--G transition, gyrochronological  relations tend to show significant scatter, likely due to weaker magnetic braking and the onset of structural evolution \citep{2016Natur.529..181V}. The measured rotation period of $\sim4.8$ days is not unusual for stars near the F--G transition and, given the large scatter in rotation--age relations in this regime, does not strongly constrain the stellar age \citep{10.1093/mnras/stv423}. In addition, a massive close companion may affect the stellar spin through tidal interactions, although the efficiency and timescale of these effects are still uncertain for F-type stars hosting massive short-period companions \citep{2014ARA&A..52..171O}}.

\subsection{TOI-2155\,b parameters}

We model TOI-2155\,b using $\tt Allesfitter$ \citep{allesfitter_paper,allesfitter_code}, a statistical tool to fit light curves and RV data and estimate the system parameters \citep{allesfitter_paper,allesfitter_code}. We employed nested sampling to sample from the posterior of the model given time-series radial velocity and photometric data, which can be provided concurrently from multiple instruments and bandpasses. $\tt Allesfitter$ wraps several other software, including models such as $\tt ELLC$ \citep{maxted2016ellc} and samplers such as $\tt DYNESTY$ \citep{speagle2020dynesty} and $\tt EMCEE$ \citep{foreman2013emcee}.

\begin{table*}
    \centering
    \caption{Derived Parameters for TOI-2155\,b from \texttt{Allesfitter}}
    \begin{tabular}{l c}
        \hline
        Parameter & Value \\
        \hline
        Host radius over semi-major axis b; $R_\star/a_\mathrm{b}$ & \change{$0.154 \pm 0.004$} \\
        Semi-major axis b over host radius; $a_\mathrm{b}/R_\star$ & \change{$6.5 \pm 0.2$} \\
        Companion radius b over semi-major axis b; $R_\mathrm{b}/a_\mathrm{b}$ & \change{$0.0091 \pm 0.0003$} \\
  
        Companion radius b; $R_\mathrm{b}$ ($\mathrm{R_{J}}$) & \change{$0.972^{+0.009}_{-0.008}$}\\
        Semi-major axis b; $a_\mathrm{b}$ ($\mathrm{R_{\odot}}$) & \change{$11.0 \pm 0.5$} \\
        Semi-major axis b; $a_\mathrm{b}$ (AU) & \change{$0.051 \pm 0.003$} \\
        Inclination b; $i_\mathrm{b}$ (deg) & \change{$83.7 \pm 0.4$} \\
        Eccentricity b; $e_\mathrm{b}$ & \change{$0.013^{+0.013}_{-0.009}$} \\
        Argument of periastron b; $w_\mathrm{b}$ (deg) & \change{$286^{+40}_{-230}$} \\
        Mass ratio b; $q_\mathrm{b}$ & \change{$0.062 \pm 0.003$} \\

        Companion mass b; $M_\mathrm{b}$ ($\mathrm{M_{J}}$) & \change{$80.6^{+1.0}_{-1.1}$} \\

        Impact parameter b; $b_\mathrm{tra;b}$ & \change{$0.72 \pm 0.02$} \\
        Total transit duration b; $T_\mathrm{tot;b}$ (h) & \change{$3.46 \pm 0.03$} \\
        Full-transit duration b; $T_\mathrm{full;b}$ (h) & \change{$2.70 \pm 0.04$} \\
        Host density from orbit b; $\rho_\mathrm{\star;b}$ (cgs) & \change{$0.37 \pm 0.03$} \\
        Companion density b; $\rho_\mathrm{b}$ (cgs) & \change{$109^{+3.1}_{-3.3}$} \\
        Companion surface gravity b; $g_\mathrm{b}$ (cgs) & \change{$(2.1 \pm 0.1) \times 10^{5}$} \\
        Equilibrium temperature b; $T_\mathrm{eq;b}$ (K) & \change{$1550 \pm 30$} \\
        Transit depth (undil.) b; $\delta_\mathrm{tr; undil; b; TESS}$ (ppt) & \change{$3.50 \pm 0.04$} \\
        Transit depth (dil.) b; $\delta_\mathrm{tr; dil; b; TESS}$ (ppt) & \change{$3.50 \pm 0.04$} \\
        Limb darkening; $u_\mathrm{1; TESS}$ & \change{$0.25^{+0.22}_{-0.16}$} \\
        Limb darkening; $u_\mathrm{2; TESS}$ & \change{$0.16^{+0.19}_{-0.26}$} \\
        Combined host density from all orbits; $\rho_\mathrm{\star; combined}$ (cgs) & \change{$0.37 \pm 0.03$} \\
        \hline
    \end{tabular}
    \label{tab:derived_params_clean}
\end{table*}
\begin{figure}
    \centering
   \includegraphics[scale=0.40]{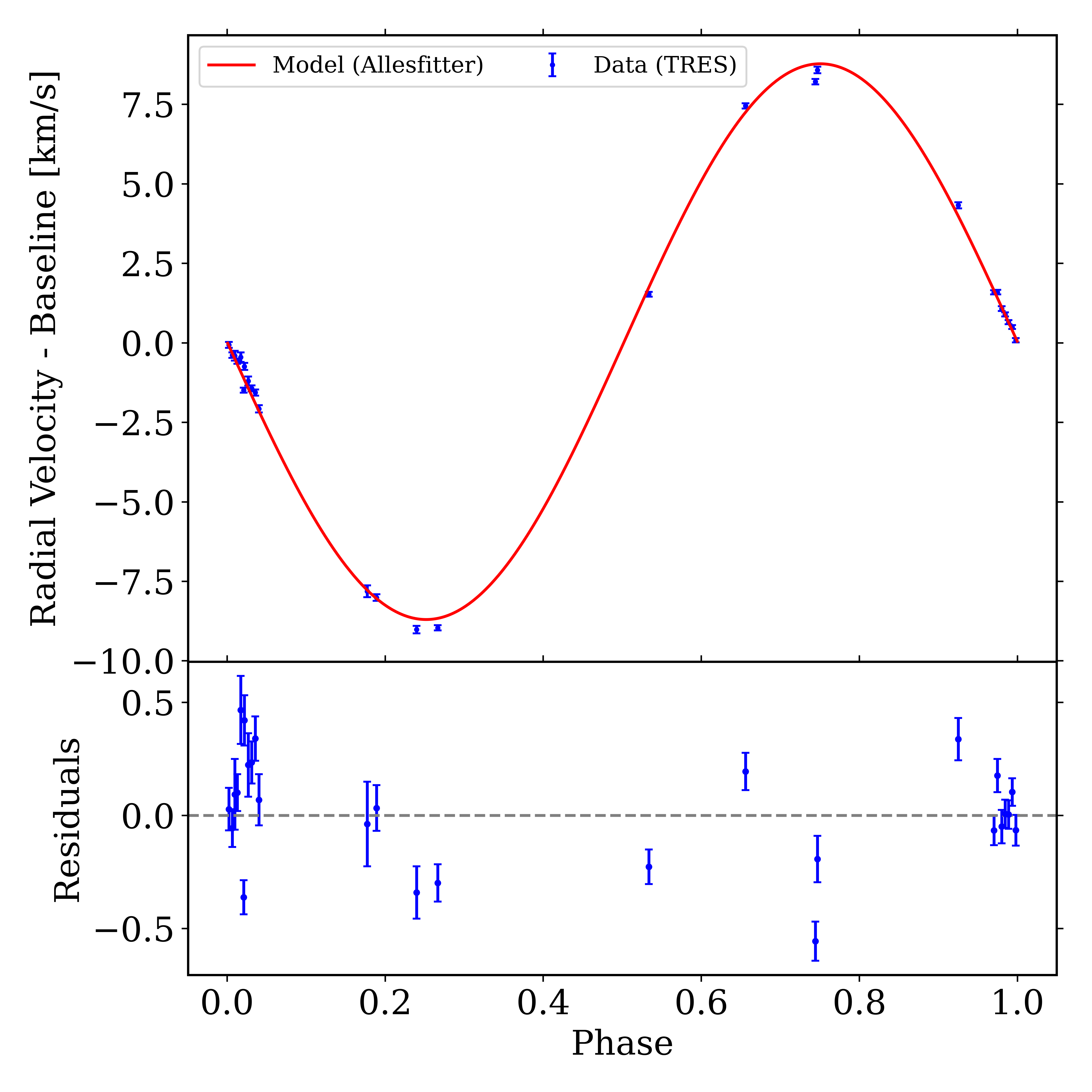}
    \caption{Blue points show the TRES multi-order relative RV measurements after subtracting the baseline component. The red curve represents the best model from $\tt Allesfitter$. The lower panel shows the residuals between the observed RVs and the full model (orbital solution + baseline).}
    \label{fig:rvplot}
\end{figure}

We used $\tt Allesfitter$ to simultaneously model the detrended TESS data and the TRES RV data. We also used it to independently model the ground-based photometric data. \change{However, the ground-based data were not included in the global modeling analysis because they do not have a sufficiently long out-of-transit baseline and do not consistently capture both ingress and egress. This limits their ability to independently constrain the transit geometry, as parameters such as the inclination $i$, impact parameter $b$, and the scaled semi-major axis $a/R_\star$ are weakly constrained and can introduce degeneracies that propagate into derived quantities such as the stellar density and the brown dwarf radius. In contrast, the TESS photometry provides continuous coverage of multiple complete transits and therefore tightly constrains the transit geometry}. The priors for all fitted parameters \change{for modelling via $\tt Allesfitter$} are listed in Table~\ref{fitted_params}.
We used the stellar radius of TOI-2155 from the SED fitting and used the estimated radius of TOI-2155\,b as listed on its ExoFOP page. Then, a rough estimate of the semimajor axis was calculated using Kepler's third law.
For the parameters R$_b$/R$_\star$ and (R$_\star$ + R$_b$)/a$_b$, we placed a uniform prior before incorporating our initial guess. \change{The eccentricity is parameterised using $\sqrt{e_b}\cos\omega_b$ and
$\sqrt{e_b}\sin\omega_b$, which are each assigned broad uniform priors
over the range $[-1,1]$. We also adopt a uniform prior on $\cos i$, allowing for a wide range of orbital geometries}.

We assign uniform prior values for the epoch. However, it is allowed to shift to avoid bias towards one data set, which would result in correlations between epoch and period. We used 500 live points and a tolerance of 0.1.
The orbital period was also fitted with a uniform prior, centered on the 3.72-day period from the TRES orbital solution. We fit the semi-amplitude with a uniform prior, which was obtained \change{from a preliminary orbital solution derived from the TRES  data}. According to the Kipping triangular quadratic method \citep{kipping2013efficient}, we assign uniform priors from 0 to 1 for the limb darkening coefficients. 

\begin{figure}
    \centering
   \includegraphics[width=\linewidth]{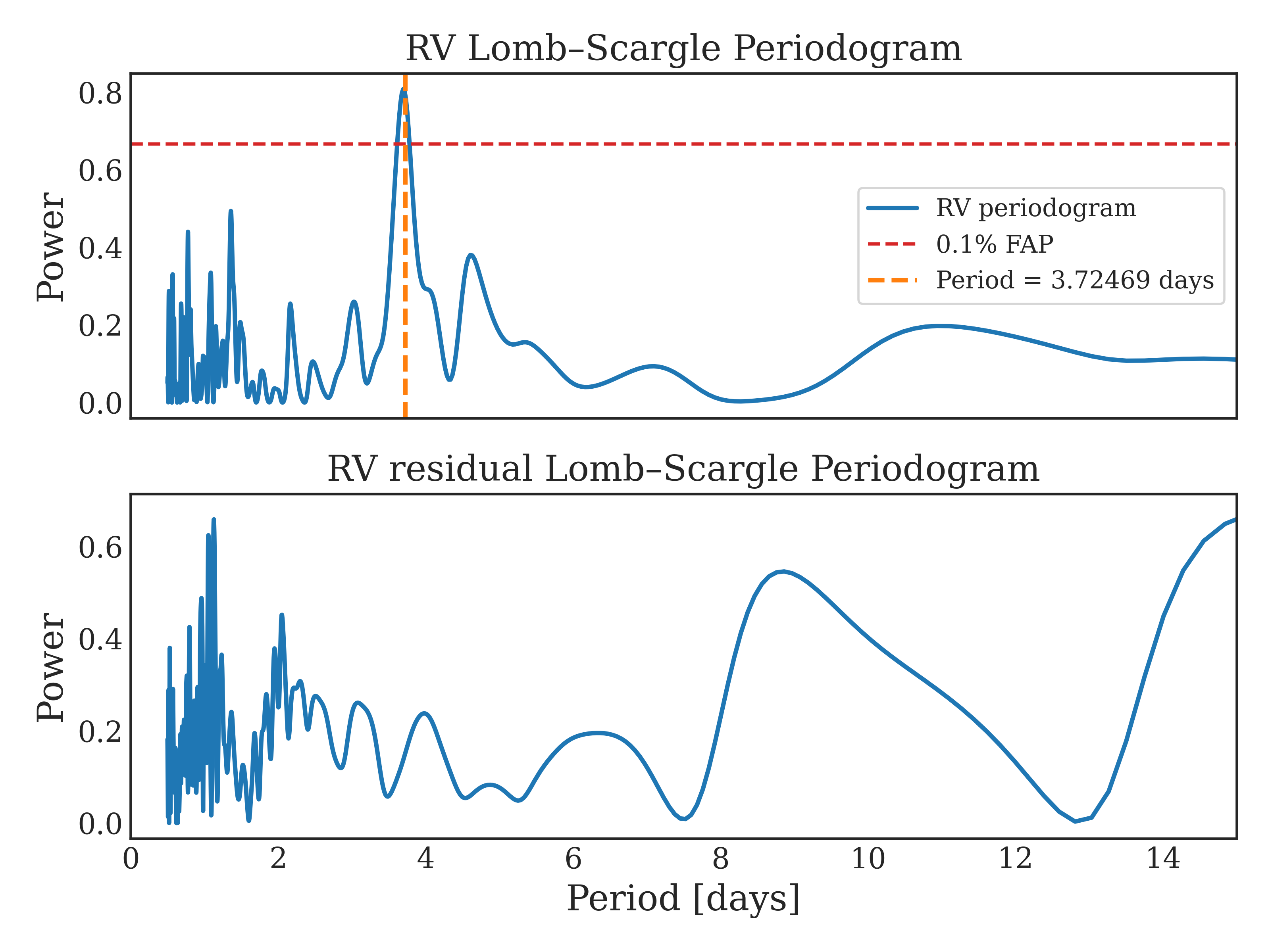}
    \caption{ \change{Lomb--Scargle periodograms of the RV measurements of TOI-2155 and of the residuals after subtracting the best-fitting orbital model. \textit{Top panel:} Periodogram of the observed RV data. The dashed red horizontal line indicates the 0.1\% false-alarm probability (FAP) threshold, and the vertical dashed line marks the orbital period of $P = 3.72469$ days derived from the RV fit. The dominant peak at this period confirms the presence of the companion. \textit{Bottom panel:} Periodogram of the RV residuals after removal of the orbital solution. No significant periodic signals are detected, indicating that the observed RV variability is well explained by the orbital motion of the companion.}}
    \label{fig:rv_periodogram}
\end{figure}

We modeled noise per dataset by fitting the logarithm of the white-noise amplitude and included an extra jitter term for TRES to capture additional scatter. Finally, we included constant baseline offsets (sample offsets), which means subtracting a constant value from each data set, for both the TESS and TRES data to correct for any zero-point differences.

We also have a total of 27 radial velocity (RV) observations, of which nine were obtained specifically during the Rossiter–McLaughlin (RM) sequence. During transit, the planet blocks part of the rotating star, causing the spectral lines to shift temporarily. This produces a short-term RV signal (the RM effect) that does not reflect the star’s true orbital motion \citep{rossiter1924detection,mclaughlin1924some,gaudi2007prospects}. To test the robustness of our mass determination to RM-induced distortions, we refitted the data after temporarily excluding the 9 RM observations. The resulting companion mass remained consistent with the full data set. Consequently, our final reported fit incorporates all 27 RV points, including the RM sequence data.

The model fitted to the TESS data and ground-based observations data is shown in Figure~\ref{fig: all_phaseplots}. Figure~\ref{fig:rvplot} illustrates the $\tt Allesfitter$ model fitted to radial velocity data from TRES. 

\change{To further assess the significance of the radial velocity signal, we computed Lomb–Scargle periodograms (Figure~\ref{fig:rv_periodogram}) of the RV data and of the residuals after subtracting the best-fitting orbital model. The RV periodogram shows a strong peak at 3.7246950 days, consistent with the orbital period derived from the RV fit. The periodogram of the residuals does not show any significant periodic signals, indicating that the observed RV variability is fully explained by the companion orbit}. The corner plot for the parameters of this fit is also in the Appendix in Figure~\ref{fig: ns_corner}. The fitted parameters are in Table~\ref{fitted_params}.

We determine that TOI-2155\,b has a mass of \change{$M_b = 80.6^{+1.0}_{-1.1} $ $ {M_J}$}, a radius of \change{ $  0.972^{+0.009}_{-0.008}$}, a density of \change{$ 109^{+3.1}_{-3.3} $} g cm$^{-3}$ with a period of \change{$P= 3.7246950 \pm{0.0000014}$}~days. The derived parameters are in Table~\ref{tab:derived_params_clean}.

\change{We additionally explored potential transit timing variations (TTVs) by fitting individual transit times for all available TESS transits. The measured timing residuals show no significant trends or periodic signals, indicating no evidence for TTVs in the TESS data.}

\section{Discussion}
\label{sec: Discussion}
The discovery of TOI-2155\,b adds another transiting \change{companion near the brown dwarf–low-mass star boundary to the growing population of objects with precisely} measured masses and radii, now exceeding 50 known systems \citep{vowell202511,henderson2024ngts,carmichael2020two,vsubjak2020toi}. \change{As outlined in Section~1, brown dwarfs are generally defined to have masses between approximately 13 and 80~$M_{\mathrm{J}}$. With a measured mass of $80.6^{+1.0}_{-1.1}\,M_{\mathrm{J}}$, TOI-2155\,b lies very close to the hydrogen-burning minimum mass limit, placing it at the transition between brown dwarfs and very low-mass stars. In this regime, objects may either sustain stable hydrogen fusion or cool and contract as brown dwarfs, depending on their interior structure and composition. Very low-mass stars such as OGLE-TR-122B \citep{pont2005planet} and EBLM~J0555$-$57Ab \citep{von2017eblm} demonstrate that objects at the bottom of the main sequence can reach densities comparable to those of massive brown dwarfs due to partially degenerate interiors. Therefore, the high density of TOI-2155\,b alone does not uniquely determine whether it is a brown dwarf or a hydrogen-burning star. TOI-2155\,b therefore lies in a regime where distinguishing between massive brown dwarfs and the lowest-mass hydrogen-burning stars remains challenging}.

\subsection{TOI-2155\,b and the Brown Dwarf Desert}

TOI-2155\,b joins the growing number of \change{massive companions} found in close-in orbits around their host star. With a mass around \change{80.6}~${M_J}$ and an orbital period of just 3.72 days, it lies in the lower part of the so-called `brown dwarf desert': a reference to the relative scarcity of BDs on short orbital periods ($P \lesssim 10$-100 days) around stars \citep{wilson2016sophie,grether2006dry, Marcy2000}. \changeblue{ Recent studies suggest that the brown dwarf desert may be more naturally described in terms of companion-to-host mass ratio rather than orbital period alone \citep{duchene2023low,zhang2026oasis}. The origin of the brown dwarf desert remains uncertain, with proposed explanations involving a combination of formation and evolutionary processes}. Those that do form might be pushed out or destroyed by interactions in the system's early stages \citep{henderson2024ngts, grieves2021populating, armitage2002}.
\begin{figure*}
    \centering
    \includegraphics[width=0.80\textwidth]{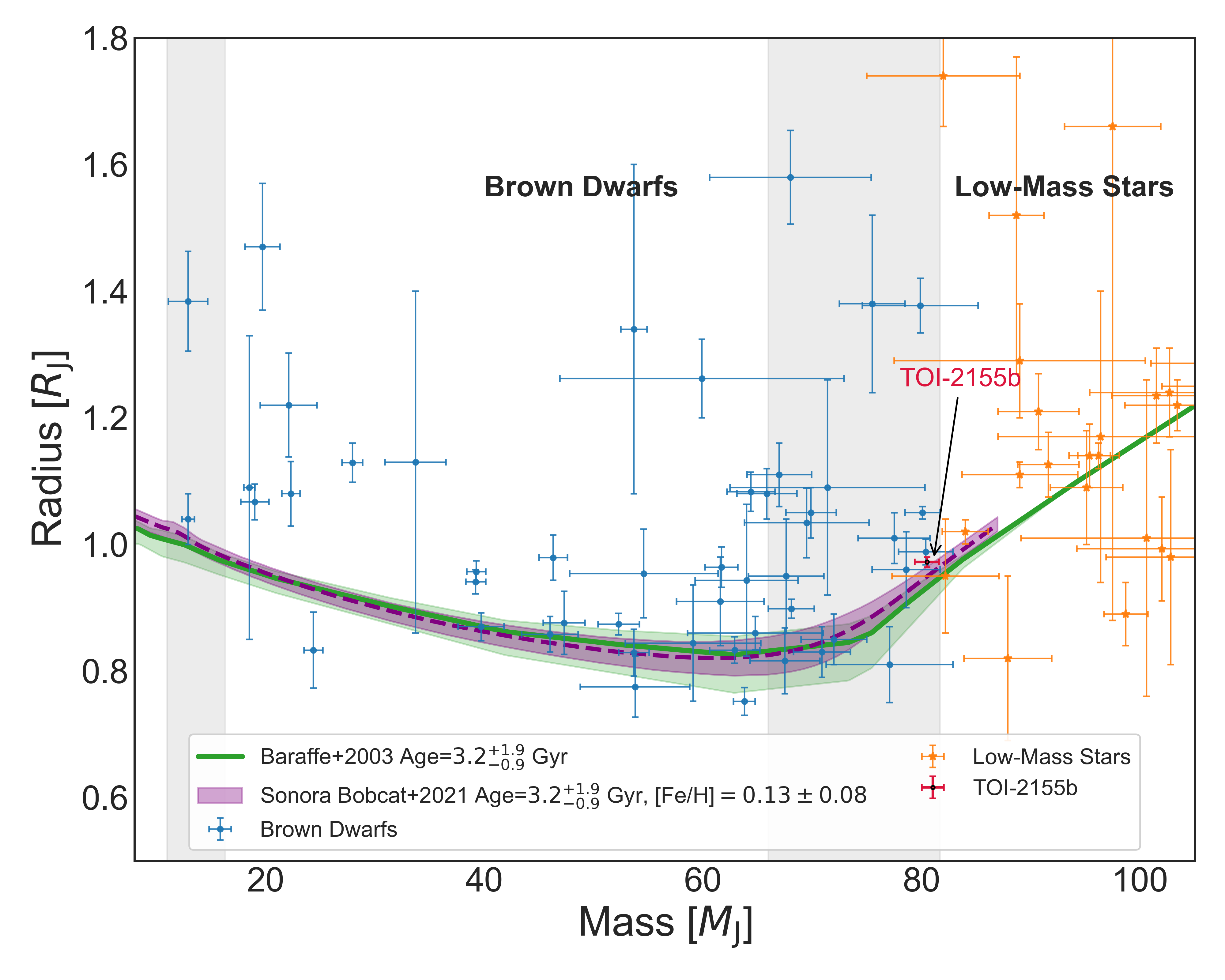} 
    \caption{\change{Mass–radius diagram for known transiting brown dwarfs (blue circles) and low-mass stars (orange stars) between $12$–$110\,M_{\mathrm{J}}$. TOI-2155\,b is highlighted in red. The green solid line shows evolutionary models from \citet{baraffe2003evolutionary} at an age of $3.2^{+1.9}_{-0.9}$ Gyr, with the shaded region indicating the corresponding age uncertainty. The purple dashed line shows Sonora Bobcat model  \citep{marley2021sonora} predictions, with uncertainties (shaded region) from age and metallicity. The grey shaded regions indicate the approximate transition mass ranges between giant planets and brown dwarfs ($\sim11$–$16.3\,M_{\mathrm{J}}$; \citealt{spiegel2011deuterium}) and between brown dwarfs and low-mass stars ($\sim66$–$81.7\,M_{\mathrm{J}}$; \citealt{morley2024sonora}). The upper boundary of  \citealt{morley2024sonora} reflects model-dependent assumptions (e.g., metallicity and cloud properties), which do not apply to TOI-2155\,b, based on the evidence we have available till now. Brown dwarfs follow a nearly flat mass–radius relation, while low-mass stars have larger radii due to sustained hydrogen fusion.}}
    \label{fig: mass_radius_diagram}
\end{figure*}

Transiting systems like TOI-2155\,b are especially valuable because the combination of transit photometry and radial velocity measurements allows us to directly measure both the mass and radius of the \change{transiting companion}, providing crucial information about its structure and evolution. Although transit detections typically limit us to exploring the parameter space of short-period BDs (usually $P_\mathrm{orb}<30$ days to detect and characterize multiple transits efficiently), they are still essential to fill out the population of \change{massive companions} with measured masses \citep{carmichael2022toi}. \changeblue{It is worth noting that recent discoveries of BDs from NASA’s TESS mission \citep[e.g.][and references therein]{refId0,vowell202511,henderson2024ngts,carmichael2022toi} are steadily increasing the number of known systems within the brown dwarf desert, while the overall scarcity of such companions remains}. 

\change{The characteristics of TOI-2155\,b make it a useful system for studying substellar companions in highly irradiated environments. Although its present-day orbital configuration may have been modified by dynamical and tidal evolution, systems like TOI-2155 still help improve our understanding of close-in transiting companions}.

\subsection{Mass–Radius–Density Relation and Radius Inflation}
With a mass of $80.6^{+1.0}_{-1.1}\,M_{\rm J}$, TOI-2155\,b lies near the
hydrogen-burning minimum mass, placing it in a regime where
distinguishing between a massive brown dwarf and a very low-mass star
is difficult.

Figure~\ref{fig: mass_radius_diagram}, adapted from Figure 12 of \cite{refId0}, shows the mass-radius distribution of BDs (blue circles) and low-mass stars (orange stars) within 12-110 $M_\mathrm{J}$. Figure~\ref{fig: mass_radius_diagram} also illustrates the isochrones for $3.2 ^{+1.9}_{-0.9}$~Gyr from the models of \citet[][solid green line and shaded green envelope]{baraffe2003evolutionary} and \citet[][dashed purple line]{marley2021sonora}. The \citet{baraffe2003evolutionary} isochrone was derived by linearly interpolating models at 1 and 5 Gyr for solar metallicity. In contrast, the \cite{marley2021sonora} isochrone was bilinearly interpolated between the Sonora Bobcat grids at ages 3 and 4 Gyr and metallicities [M/H] = 0.0 and +0.5 to match the system’s age of 3.2 Gyr and metallicity [M/H]=0.13. \change{In Figure~\ref{fig: mass_radius_diagram}, TOI-2155\,b (red circle) lies slightly above the central 3.2 Gyr isochrones from both \citet{marley2021sonora} and \citet{baraffe2003evolutionary}. However, when the uncertainties in age ($3.2^{+1.9}_{-0.9}$ Gyr) and metallicity ([M/H]$=0.13\pm0.08$) are considered, the measured radius remains consistent with the theoretical uncertainty envelopes of the evolutionary models. This indicates that the radius of TOI-2155\,b agrees with evolutionary models when uncertainties are taken into account}.


The measured radius of TOI-2155\,b is slightly larger than the central Sonora model prediction (by $\sim1.6\%$). Still, it remains consistent with the evolutionary tracks once the uncertainties in age and metallicity are taken into account. \changegreen{Stellar irradiation has been suggested to influence the radii of strongly irradiated massive transiting brown dwarfs \citep{mukherjee2025}. However, this effect is expected to be more modest for high-mass brown dwarfs because of their stronger surface gravity.
Separately, magnetic activity and the associated inhibition of convection have been proposed as mechanisms for radius inflation in low-mass stars \citep{10.1007/978-3-642-11250-8_114,2013ApJ...779..183F}.  Similar effects, including reduced convective efficiency and magnetic spot coverage have also been suggested as mechanisms influencing the evolution and radii of brown dwarfs \citep{chabrier2007}. However, the extent to which magnetic activity and convection inhibition contribute to radius inflation in massive transiting companions such as TOI-2155\,b remains less well established, although the high internal densities and pressures of such objects may reduce the influence of magnetic stresses on convective flows \citep{2010ApJ...713.1249M}, while increasing electron degeneracy may make their radii less sensitive to changes in convective heat transport.}

\begin{figure*}
      \centering
    \includegraphics[width=0.85\textwidth]{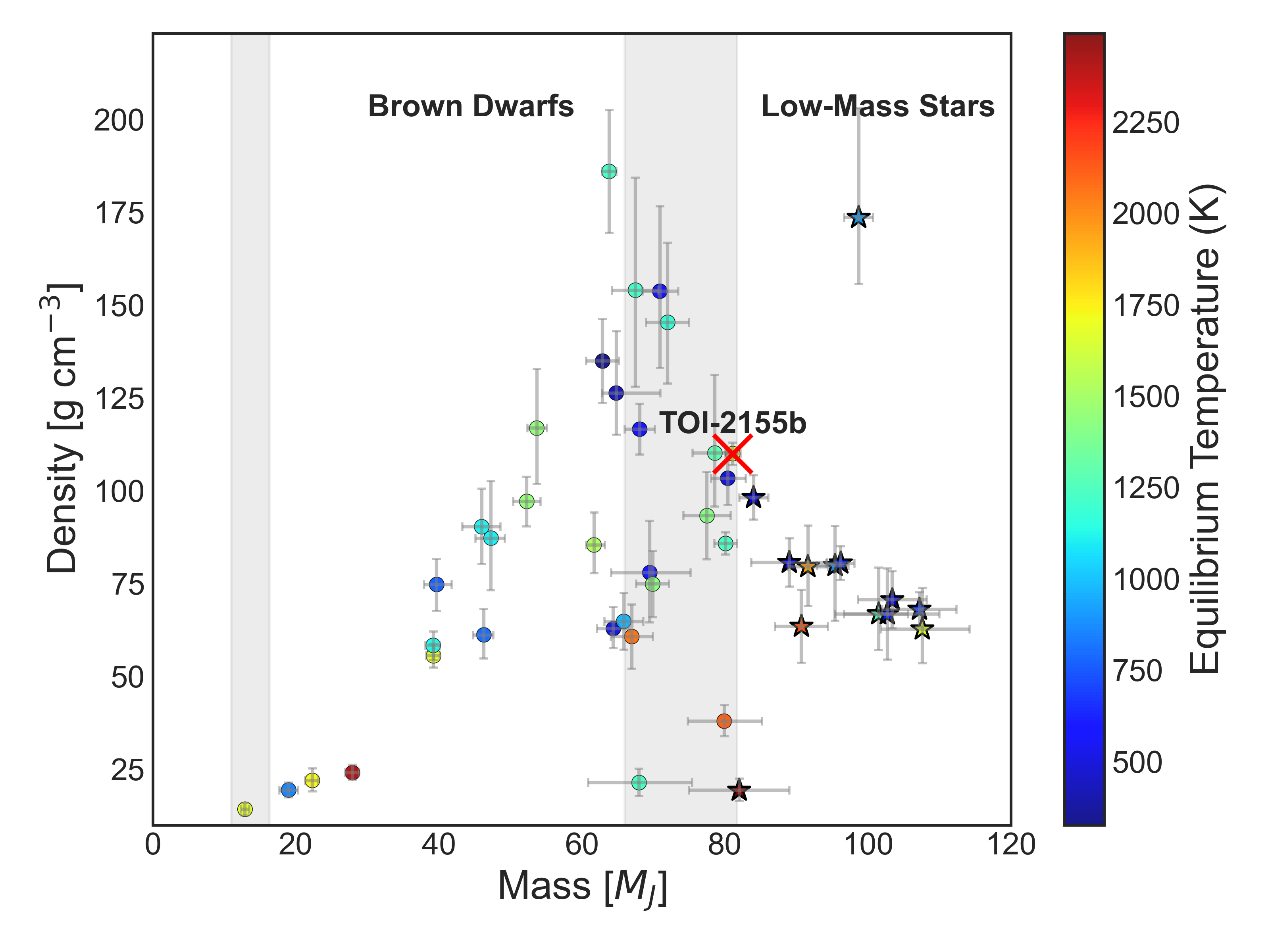}
\caption{\changeblue{Mass–density distribution of transiting brown dwarfs (BDs) and low-mass stars, depicting systems with precision \(>20\%\) in both mass and density \citep[e.g.,][]{persson2019greening}. Brown dwarfs are shown as circles, color-coded by equilibrium temperature ($T_{\mathrm{eq}}$) and with marker size scaled by radius, while low-mass stars are shown as star symbols with the same colour coding. The shaded regions indicate the approximate transition regimes corresponding to the deuterium-burning limit ($\sim11$–$16.3\,M_{\mathrm{J}}$; \citealt{spiegel2011deuterium}) and the hydrogen-burning boundary ($\sim66$–$81.7\,M_{\mathrm{J}}$; \citealt{morley2024sonora}). The distribution exhibits a turnover in density around $\sim60$--$70\,M_{\mathrm{J}}$. TOI-2155\,b is highlighted with a red ‘X’ over its marker, annotated for emphasis. }}
\label{fig: Mass_vs_density}
\end{figure*}

The high density (see Figure~\ref{fig: Mass_vs_density}) of TOI-2155\,b suggests that its interior is likely dominated by \change{electron degeneracy pressure rather than thermal pressure. Such conditions are expected for objects near the hydrogen-burning minimum mass, where degeneracy begins to play a significant role in supporting the interior structure}
\citep{Chabrier_2000,burrows1997nongray}. This implies that the object has cooled and contracted significantly over its lifetime.


\changeblue{Near the upper end of the BD to low-mass star transitionary mass regime, the position of TOI-2155\,b is highlighted in the mass–density diagram of Figure~\ref{fig: Mass_vs_density}.
The distribution shows an increase in density with mass up to $\sim60$--$70\,M_{\rm J}$, where it reaches a maximum. At higher masses, the densities are generally lower than this peak and consistent with the expected turnover in the mass–density relation. TOI-2155\,b, with a mass of $80.6^{+1.0}_{-1.1}\,M_{\rm J}$ and a density of $\sim109\,\mathrm{g\,cm^{-3}}$, lies close to this turnover and slightly beyond the peak region. However, objects in the 60–70 $M_{\rm J}$ range show a large spread in densities, with no clear correlation with equilibrium temperature. Transiting systems such as TOI-2155\,b further highlight this diversity, rather than supporting a clear density turnover driven purely by mass. 
Objects in this regime, near the hydrogen-burning limit, are particularly valuable for constraining the mass-density boundary between massive brown dwarfs and very low-mass stars, where degeneracy pressure and nuclear burning both play a role \citep{hatzes2015definition,spiegel2011deuterium}. Characterizing such high-mass companions in this transition region, therefore, can help us define the difference between BDs and low-mass stars.}

\subsection{Orbital Characteristics, Eccentricity, and Tidal Evolution}

\begin{figure}
    \centering
    \includegraphics[width=\linewidth]{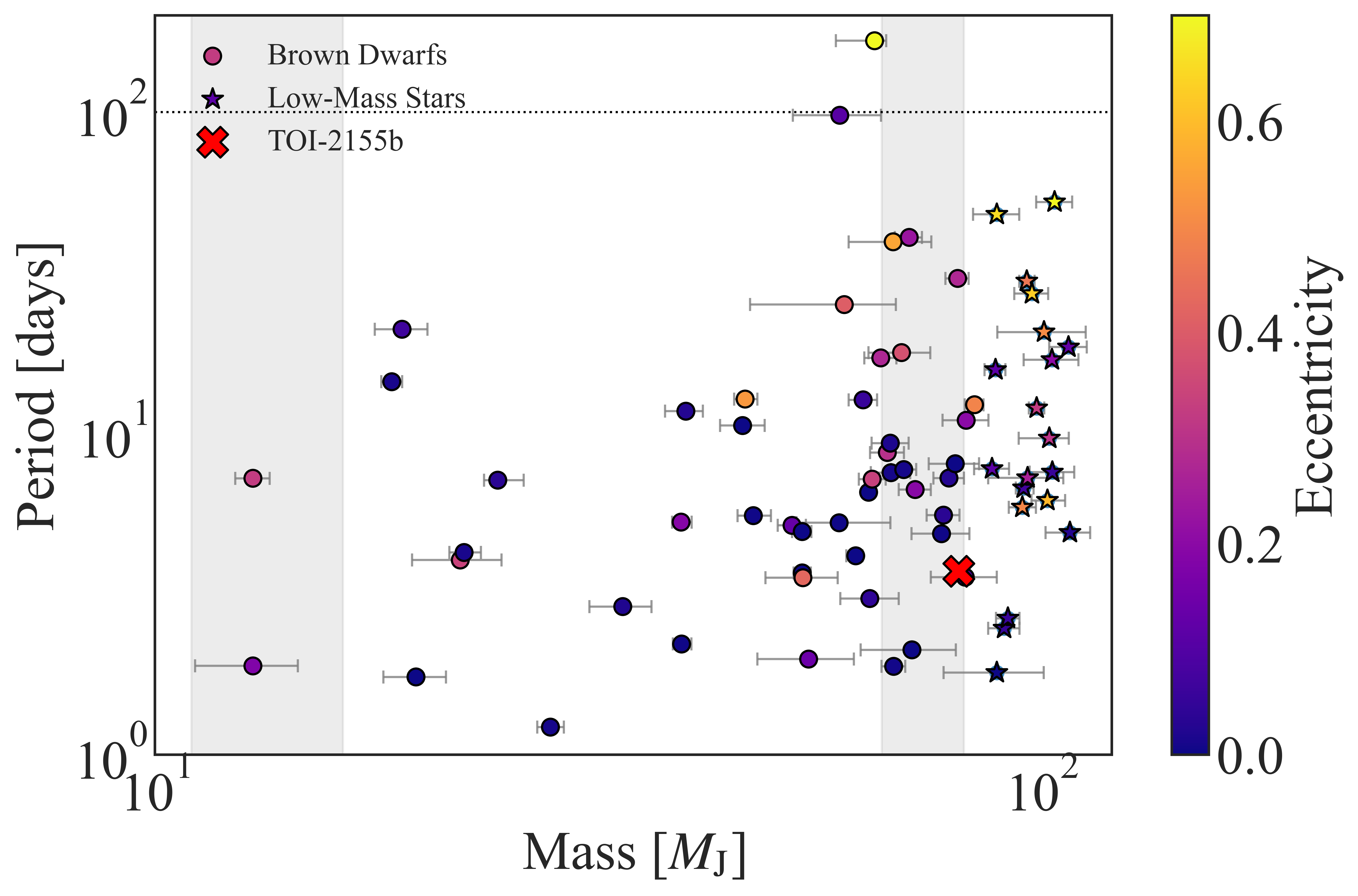}
    \caption{Orbital period-Mass diagram for transiting BDs (circles) and low-mass stars (star symbols), color-coded by orbital eccentricity. TOI-2155\,b is highlighted with a red X. \change{The shaded regions indicate the approximate transition regimes corresponding to the deuterium-burning limit ($\sim$11–16.3~$M_{\mathrm{J}}$) and the hydrogen-burning boundary ($\sim$66–81.7~$M_{\mathrm{J}}$)}, while the dotted horizontal line at 100 days illustrates the transition between short- and long-period systems. A trend emerges where longer-period objects tend to exhibit higher eccentricities, consistent with less efficient tidal circularization. In contrast, short-period BDs like TOI-2155\,b typically occupy the low-eccentricity regime. High-mass BDs exhibit a broad range of eccentricities, underscoring the diversity of their dynamical histories and potential formation pathways. }
    \label{fig:mass_period_ecc}
\end{figure}

To further explore the dynamical properties of the population of \change{transiting companions} found in orbit around stars, we present an orbital period-mass diagram color-coded by orbital eccentricity in Figure~\ref{fig:mass_period_ecc}. This diagram highlights a clear trend: systems with longer orbital periods tend to exhibit higher and more diverse eccentricities, consistent with incomplete tidal circularization. In contrast, short-period systems such as TOI-2155\,b generally show low eccentricities, supporting the interpretation that tidal forces have acted efficiently on Gyr timescales to circularize their orbits \citep{psaridi2022three}. \changeblue{TOI-2155\,b has a low eccentricity ($e = 0.013^{+0.013}_{-0.009}$), consistent with its short orbital period (3.72~days) and the expected tidal circularization over its $\sim$3.2~Gyr lifetime.}

 We investigated the tidal evolution of the system \change{using the equilibrium tide formalism with a constant tidal quality factor following \citet{Jackson_2008}, estimating the orbital circularization timescale for the system. Using the small-eccentricity approximation, appropriate for the nearly circular orbit ($e_b \approx 0.013$), this formulation provides the e-folding timescale for eccentricity damping rather than the time required to reach exactly zero eccentricity. Adopting conservative tidal quality factors of $Q_\star = 10^7$ for the host star and $Q_{\rm BD} = 10^5$ for the brown dwarf \citep{henderson2024ngts}, we obtain a circularization timescale of $\tau = 0.33$ Gyr, which is significantly shorter than the system age of $\sim3.2$ Gyr. These timescales, however, can significantly change for short-period planets/BDs with the inclusion of energy dissipation via excited inertial waves in stellar convective regions \citep{Alvarado-Montes2019,Alvarado-Montes2021}. This will be explored in a future study.}
\begin{figure}
\centering
\includegraphics[width=\linewidth]{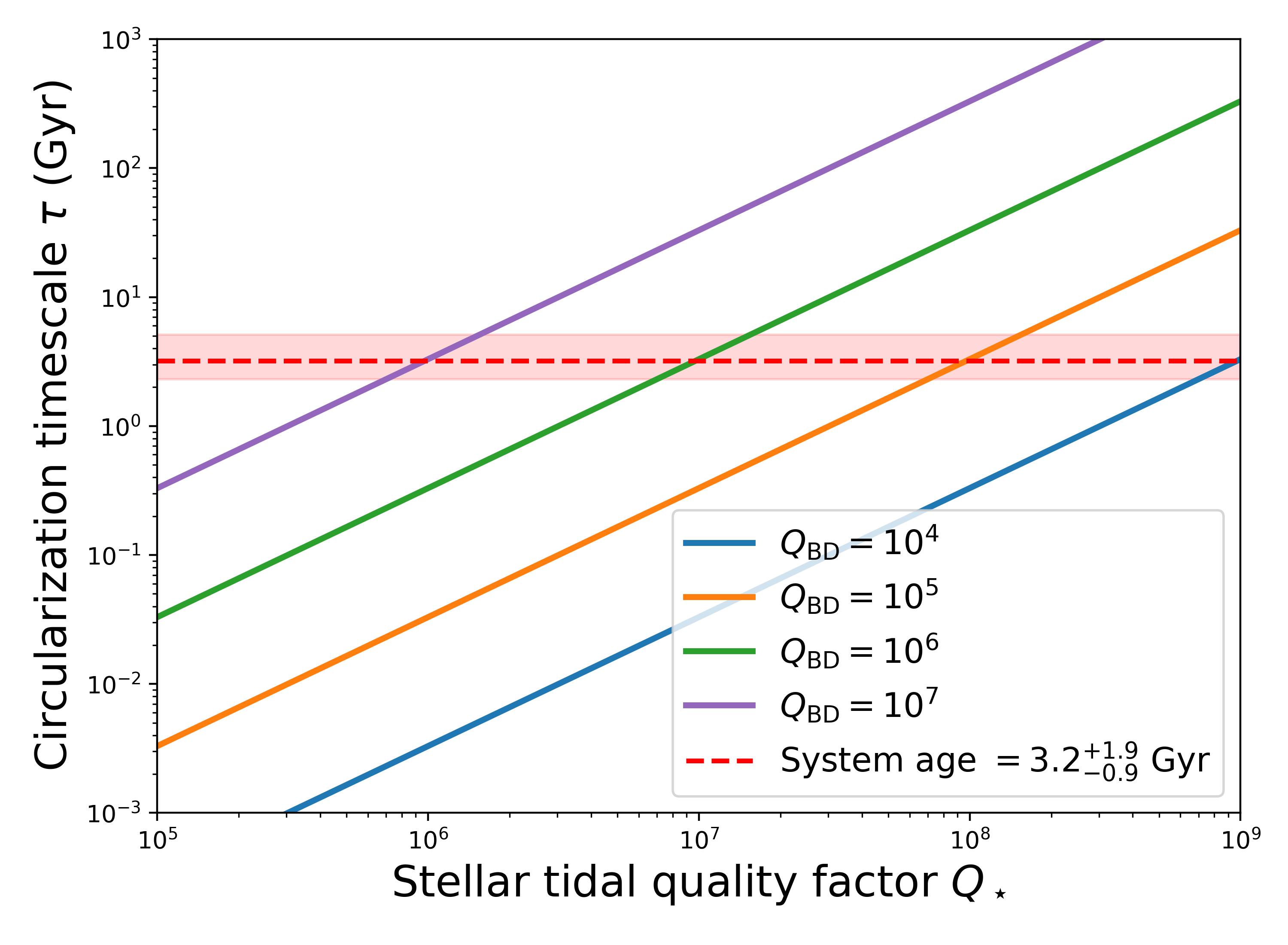}
\caption{\change{Circularization timescale $\tau$ as a function of the stellar tidal quality factor $Q_\star$ calculated using the equilibrium tide formalism of \citet{Jackson_2008}. The colored curves correspond to different assumed brown dwarf tidal quality factors $Q_{\rm BD}$. The dashed red line marks the system age of $3.2^{+1.9}_{-0.9}$\,Gyr, with the shaded region indicating its uncertainty. For typical tidal quality factors, the calculated
timescales lie below the system age.}}
\label{fig:tidal_timescale}
\end{figure}
\change{To illustrate the dependence of the circularization timescale on tidal dissipation efficiency, Figure~\ref{fig:tidal_timescale} shows $\tau$ as a function of the stellar tidal quality factor $Q_\star$ for different assumed values of $Q_{\rm BD}$. For \changeblue{commonly adopted tidal quality factors ($Q_{\rm BD} \sim 10^5$–$10^7$)}, the circularization timescale remains shorter than the system age of $3.2^{+1.9}_{-0.9}$ Gyr, suggesting that tidal dissipation could have efficiently damped the orbital eccentricity. Given the large mass of TOI-2155\,b, the breaking of internal gravity waves in stellar radiative regions \citep{Barker2020} can also contribute to tidal evolution, as previously explored for large Jupiter-like planets orbiting very close to their stars \citep{Alvarado-Montes2025} and also around M dwarfs \citep{Alvarado-Montes2022,Lin2023,Gan2023}. Exploring these effects is out of the scope of this work.}

\subsection{Age, Formation Pathway and Mass Ratio}
\change{TOI-2155\,b lies near the hydrogen-burning minimum mass}, with a system mass ratio of $0.061_{-0.003}^{+0.003}$, exceeding the typical planet–star boundary of approximately 0.01 \citep{poleski2017companion,hatzes2015definition}. 
Its high mass ratio places it closer to the star-star regime than the planet–star regime, suggesting a stellar-like formation pathway, such as gravitational collapse or fragmentation of molecular clouds, \change{or formation via disc fragmentation} \citep{bowler2020population}. Most BDs likely form through such processes, with only a few exceptions for companions orbiting very massive host stars \citep{duchene2023low}.

Finally, given TOI-2155\,b's mature age ($\approx3.2$ Gyr) and lack of detectable lithium absorption, it is expected to have cooled and contracted significantly. 

\section{Conclusion}
\label{sec:conclusion}

\change{In conclusion, TOI-2155\,b stands out as a massive transiting companion
lying near the hydrogen-burning minimum mass, placing it close to the
boundary between brown dwarfs and very low-mass stars. With a precisely
measured mass of $80.6^{+1.0}_{-1.1}\,M_{\rm J}$ and a radius of
$0.972^{+0.009}_{-0.008}\,R_{\rm J}$, TOI-2155\,b occupies a regime
where the distinction between stellar and substellar objects becomes
particularly challenging}.

\change{The measured radius of TOI-2155\,b is broadly consistent with
theoretical evolutionary models once uncertainties in the system age
($3.2^{+1.9}_{-0.9}$\,Gyr) and metallicity are taken into account.
In particular, the observed radius agrees with predictions from
substellar evolutionary models such as those of
\citet{baraffe2003evolutionary} and the Sonora Bobcat models
\citep{marley2021sonora} within their expected uncertainty ranges. Although we find no robust evidence for radius inflation in this case,
several recently discovered massive companions near the brown dwarf–very low-mass star boundary have shown inflated radii
(e.g., \citealt{schmidt2023,grieves2021populating,casewell2020nltt5306b})}.

\change{Very low-mass stars such as OGLE-TR-122B \citep{pont2005planet}
and EBLM~J0555$-$57Ab \citep{von2017eblm} demonstrate
that objects at the bottom of the main sequence can exhibit densities
comparable to those of massive brown dwarfs due to partially degenerate interiors. Therefore, despite the precise measurements of the mass, radius, density, and age of TOI-2155\,b, and comparisons with
theoretical evolutionary models, we cannot unambiguously determine whether TOI-2155\,b is a massive brown dwarf or an extremely
low-mass hydrogen-burning star}.

\change{TOI-2155\,b's short orbital period of 3.72 days and nearly circular
orbit ($e = 0.013^{+0.013}_{-0.009}$) suggest that tidal dissipation
may have played an important role in shaping the system's present
orbital configuration \citep{Jackson_2008}. In addition, the system
mass ratio ($q \approx 0.061$) is significantly larger than the
typical planet–star boundary ($q \sim 0.01$), supporting a formation
pathway more similar to stellar companions than to giant planets
\citep{hatzes2015definition,poleski2017companion}}.

\change{Objects in this mass regime provide valuable empirical constraints
on the transition between stellar and substellar objects. Systems
such as TOI-2155\,b, therefore, represent important benchmarks for
testing evolutionary models near the hydrogen-burning limit and
for improving our understanding of the formation and evolution of
companions at the boundary between brown dwarfs and very low-mass stars}.

\section{acknowledgments}
M.R.A. is supported by a postgraduate research scholarship from the University of Sydney.

T.D. acknowledges support from the McDonnell Center for the Space Sciences at Washington University in St. Louis.

J.A.A.M. is supported via the Macquarie University Research Fellowship (MQRF).

Funding for the TESS mission is provided by NASA’s Science Mission Directorate. This paper includes data collected by the TESS mission, which are publicly available from the Mikulski Archive for Space Telescopes (MAST) \change{at the Space Telescope Science Insti-
tute. The specific observations analyzed can be accessed via \dataset[doi:10.17909/2fd7-4p25]{https://doi.org/10.17909/2fd7-4p25}}.  We acknowledge the use of public TESS data from pipelines at the TESS Science Office and at the TESS Science Processing Operations Center. Resources supporting this work were provided by the NASA High-End Computing (HEC) Program through the NASA Advanced Supercomputing (NAS) Division at Ames Research Center for the production of the SPOC data products.


This work makes use of observations from the LCOGT network. Part of the LCOGT telescope time was granted by NOIRLab through the Mid-Scale Innovations Program (MSIP). MSIP is funded by NSF.


This research has made use of the Exoplanet Follow-up Observation Program (ExoFOP; DOI: 10.26134/ExoFOP5) website, which is operated by the California Institute of Technology, under contract with the National Aeronautics and Space Administration under the Exoplanet Exploration Program. \changegreen{ This research made use of the Uzay compute node at the MIT Kavli Institute for Astrophysics and Space Research.}
\changeblue{This work made use of \texttt{TESS-cont} (\url{https://github.com/castro-gzlz/TESS-cont}), which also made use of \texttt{tpfplotter} \citep{2020A&A...635A.128A} and \texttt{TESS-PRF} \citep{2022ascl.soft07008B}}.

We thank Juan Carlos Morales and K. R. Sreenivas for useful discussions and suggestions to improve this paper. \changegreen{We acknowledge and pay respect to the traditional owners of the land on which the University of Sydney and Macquarie University are situated, upon whose unceded, sovereign, ancestral lands we work. We pay respects to their Ancestors and descendants, who continue cultural and spiritual connections to Country.  We are grateful to the Australian public for enabling this science.}

\section{Data availability}
The TESS data is available through the MAST (Mikulski Archive for Space Telescopes) portal. TFOP data for TOI-2155\,b may be accessed at \href{https://exofop.ipac.caltech.edu/tess/target.php?id=461591646}{ExoFOP-TESS}. The TRES and ground-based photometric data in this article will be shared with the associated authors upon request.

\software{$\tt Astropy$ \citep{2018AJ....156..123A, 2013A&A...558A..33A}}, {$\tt Allesfitte$r}\citep{allesfitter_code,allesfitter_paper}

\bibliography{library}

\bibliographystyle{aasjournal}
\begin{figure*}
    \centering
    \includegraphics[width=\linewidth]{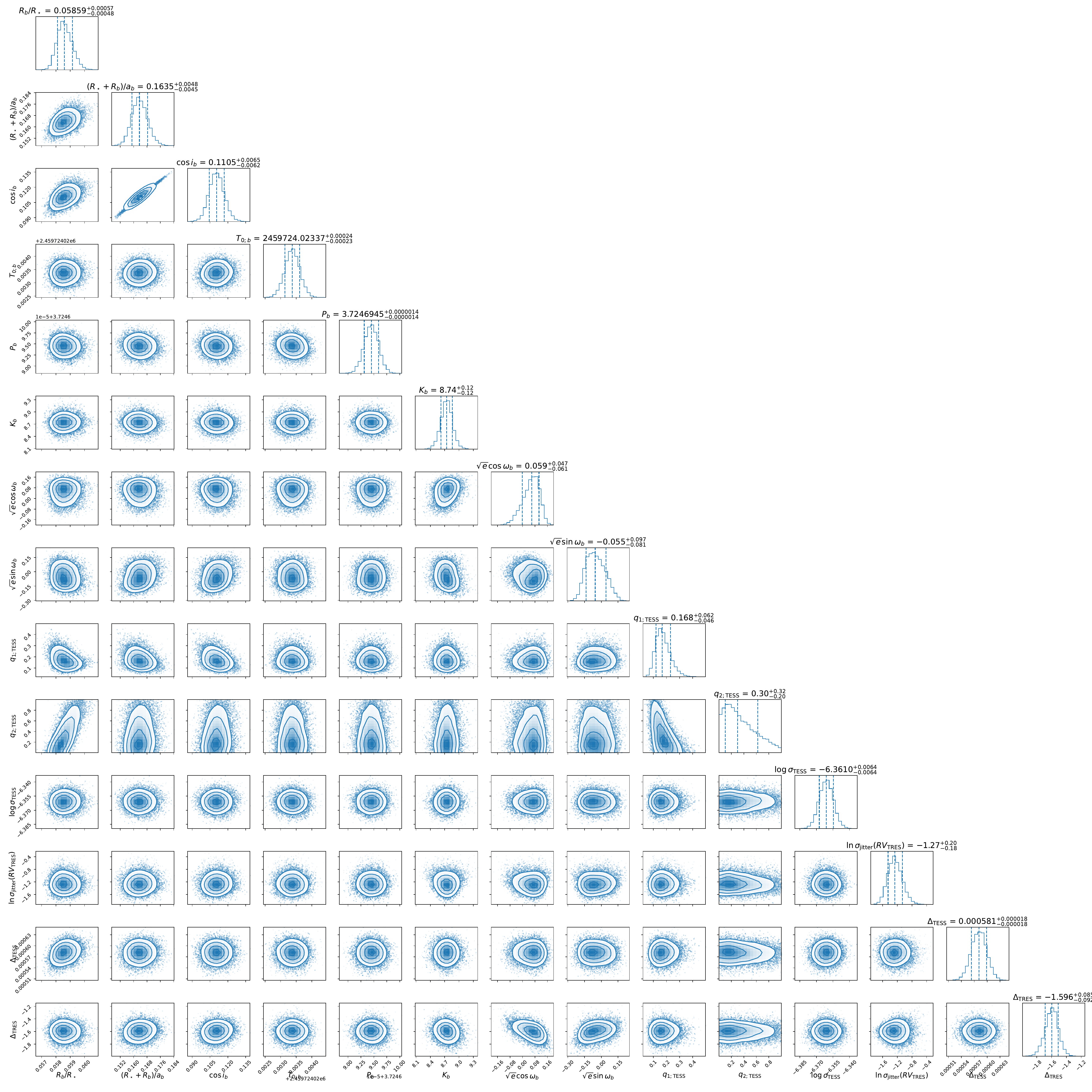}
    \caption{ The corner plot from the nested sampling fit of TOI-2155\,b.}
    \label{fig: ns_corner}
\end{figure*}

\end{document}